\documentclass{jpp}
\usepackage{graphicx}
\usepackage{epstopdf, epsfig}
\usepackage{hyperref}
\DeclareMathAlphabet\mathbfcal{OMS}{cmsy}{b}{n}

\shorttitle{Benchmarking magnetized coupling}
\shortauthor{Y. Shi}

\title{Benchmarking magnetized three-wave coupling for laser backscattering: Analytic solutions and kinetic simulations}

\author{Yuan Shi \corresp{\email{shi9@llnl.gov}}}

\affiliation{Lawrence Livermore National Laboratory, Livermore, California 94550, USA}

\begin{document}

\maketitle

\begin{abstract}
Understanding magnetized laser-plasma interactions is important for controlling magneto-inertial fusion experiments and developing magnetically assisted radiation and particle sources. In the long-pulse regime, interactions are dominated by coherent three-wave interactions, whose nonlinear coupling coefficients become known only recently when waves propagate at oblique angles with the magnetic field. 
In this paper, backscattering coupling coefficients predicted by warm-fluid theory is benchmarked using particle-in-cell simulations in one spatial dimension, and excellent agreements are found for a wide range of plasma temperatures, magnetic field strengths, and laser propagation angles, when the interactions are mediated by electron-dominant hybrid waves.
Systematic comparisons between theory and simulations are made possible by a rigorous protocol: On the theory side, the initial boundary value problem of linearized three-wave equations is solved, and the transient-time solutions allow effects of growth and damping to be distinguished. On the simulation side, parameters are carefully chosen and calibration runs are performed to ensure that stimulated runs are well controlled. Fitting simulation data to analytical solutions yields numerical growth rates, which match theory predictions within error bars. 
Although warm-fluid theory is found to be valid for a wide parameter range, genuine kinetic effects have also been observed.

\end{abstract}

\section{Introduction}
Laser-plasma interactions (LPI) are usually studied without a background magnetic field, partly because the relevant field strengths are over hundreds of teslas that are difficult to attain, and partly because cyclotron motion significantly complicates physical processes making the interactions difficult to understand. 
However, in recent experiments where a seed magnetic field is imposed to enhance thermal and particle confinements in laser-driven inertial fusion experiments \citep{chang2011fusion,hohenberger2012inertial}, understanding magnetized LPI (MagLPI) have become necessary. In experiments that are designed using codes that incorporate magnetization effects on hydrodynamics but not including magnetization effects on LPI, the observed hot spot shape is more elongated along the magnetic field than expected \citep{Moody22}. Subsequent experiments using manually adjusted laser drive manage to restore the symmetry, suggesting that MagLPI is one of the likely causes of the discrepancy.

Strong magnetic fields directly affect LPI in addition to changing plasma conditions. In indirect-drive experiments, external coils apply a seed magnetic field of \mbox{$B_0\approx30$ T}, which is not particularly strong. However, during the laser drive, expanding plasmas from the hohlraum wall compresses the magnetic flux near the laser entrance hole, causing the magnetic field there to be amplified to $100$-T level \citep{strozzi2015imposed}. Moreover, Biermann-battery fields near the laser spot and flux compression due to the imploding fuel lead to even larger fields at kT-level \citep{knauer2010compressing,sio2021diagnosing}. 
When \mbox{$B_0\sim10^2$ T}, electron cyclotron frequency becomes comparable to the frequency of sound waves that mediate Brillouin scattering and crossbeam laser energy transfer.
Moreover, when \mbox{$B_0\sim10^3$ T}, electron cyclotron frequency becomes comparable to plasma frequency, leading to modifications of the Langmuir wave that mediates Raman scattering and two-plasmon decay.  
The ability to explain and predict magnetized version of these commonly encountered long-pulse LPI processes rely on basic understandings of MagLPI that we have just begun to acquire.

While basic facts about unmagnetized LPI, such as the linear growth rates of Raman and Brillouin scatterings, are well known from decades of theoretical, numerical, and experimental studies, simple facts about MagLPI are poorly understood. 
The two exceptions are when waves propagate either perpendicular or parallel to the background magnetic field. 
Perpendicular propagation is particularly relevant for magnetic confinement fusion, where strong radio-frequency pump waves are used for plasma heating and current drive. Using antenna mounted on vacuum chamber walls, the waves are launched nearly perpendicular to the magnetic field. In this geometry, the extraordinary (X) pump can decay to upper-hybrid (UH) and lower-hybrid (LH) waves \citep{grebogi1980brillouin, hansen2017parametric}, as well as couple with Bernstein waves \citep{Platzman68,Stenflo81}.
The other special geometry is when waves propagate nearly parallel to the magnetic field, which is particularly relevant for astrophysical type plasmas, where the pump wave is an Alfv\'en wave whose frequency is below cyclotron frequencies. In this case, the pump can couple with sound waves and other Alfv\'en type waves \citep{hasegawa1976kinetic, wong1986parametric}, and the coupling is known also at oblique angles \citep{vinas1991parametric}.
In the context of LPI, the pump waves are lasers, whose frequencies are typically higher than electron-cyclotron frequency. Most studies of MagLPI also focus on special geometry where theories are greatly simplified. In the perpendicular geometry, X-wave pump lasers undergo backscattering \citep{paknezhad2016effect,shi2017laser}, forward scattering \citep{hassoon2009stimulated,babu2021stimulated}, second harmonics generation \citep{jha2007second}, and THz radiation generation \citep{varshney2015strong}. In the parallel geometry, the electrostatic waves are unmagnetized but the right-handed (R) and left-handed (L) circularly polarized electromagnetic waves become nondegenerate, changing the coupling for Raman and Brillouin type scatterings \citep{sjolund1967non,laham1998effects}. 
In more general geometry, where waves propagate at oblique angles with respect to the magnetic field, cyclotron motion makes theoretical analysis significantly more complicated. Although theories exist \citep{larsson1973three,liu1986parametric,stenflo1994resonant,brodin2012three}, the coupling coefficients are given by cumbersome formulas that are general but rarely evaluated. Physical understanding of underlying processes are largely lacking until recently when more convenient formulas are obtained and evaluated \citep{Shi17,Shi19}, leading to  pictures of MagLPI that are both intuitive and quantitative \citep{shi2018laser,shi2021plasma}. However, it remains unclear how accurate are these formulas and to what extent are they applicable. 
Because lasers propagate at oblique angles in inertial fusion experiments, it is especially important to known whether the predicted couplings are correct beyond special angles.

To benchmark analytical formulas, kinetic simulations have been used but systematic comparisons are difficult. When waves propagate perpendicular \citep{boyd1985kinetic,jia2017kinetic} or parallel \citep{li2020boosting,li2021boosting} to the magnetic field, qualitative agreements between theories and simulations have been found. Moreover, at oblique angles, it is observed that even in regimes where kinetic effects are expected to be important, coupling coefficients predicted by warm-fluid theory are indictive of kinetic simulation results \citep{edwards2019laser,Manzo22}. 
However, systematic comparisons between theory and simulations of MagLPI have not been made, which is the goal of this paper. Making the comparisons rigorous is difficult due to three reasons. 
First, nonlinear coupling leads to wave growth, but the effect of growth is mixed with damping in kinetic simulations. In the absence of collision, magnetized plasma waves are still damped collisionlessly, whose rate is difficult to calculate because cyclotron motion mixes with trapping motion and particle trajectories are chaotic in general. Even in the simple perpendicular geometry, the two limiting cases, where trapping motion dominates cyclotron motion \citep{sagdeev1973influence,dawson1983damping} or vice versa \citep{karney1978stochastic,karney1979stochastic}, have drastically different behavior, and the intermediate regimes are far less understood. Since a larger damping can offset a larger growth, their effects need to be separated before coupling coefficients can be constrained.
Second, the coupling coefficients derived in theories are specific for eigenmodes but launching eigenmodes in kinetic simulations is not straightforward when waves propagate at oblique angles. In previously simulations, the pump laser is launched with simple linear or circular polarizations from the vacuum region. Upon entering the magnetized plasma, which is a birefringent medium, the laser excites both eigenmodes, which are elliptically polarized at oblique angles. Since nonlinear couplings are different depending on the laser polarization \citep{shi2019amplification}, exciting multiple modes does not allow a clean comparison.
Finally, additional processes can occur in kinetic simulations, making it difficult to isolate the process of interest. For example, the pump laser can undergo spontaneous scattering into other modes. This problem is particularly severe in particle-in-cell (PIC) simulations, where Monte Carlo sampling noise cause unphysically large spontaneous scattering. As another problem, collisionless damping causes the distribution function to evolve on a time scale that is often comparable to wave growth, which is an issue for both PIC and Vlasov simulations. Since coupling coefficients depend on the distribution function, its time evolution complicates the comparison between theories and simulations.

In this paper, one dimensional PIC simulations are used to benchmark coupling coefficients predicted by warm-fluid theory, and excellent agreements between simulations and theory are achieved using a protocol that enables quantitative comparisons. 
First, to distinguish effects of damping from growth, analytical solutions of the linearized three-wave equation are derived in Sec.~\ref{Sec:Analytic} and are used to fit simulations data. Building up solutions from initial value problem to boundary value problem, the transient-time spatial profiles in the backscattering problem, where a seed laser is amplified by a counter propagating pump laser, allows damping and growth rates to be constrained separately. 
Second, to make numerical results directly comparable to theory, calibration steps are performed for PIC simulations, which are described in Sec.~\ref{Sec:Simulations}. 
To ensure that a single eigenmode is excited, the pump and seed laser polarizations are calibrated. 
To account for laser reflections from plasma-vacuum boundaries, laser transmissions are calibrated. 
To separate pump and seed from raw data and extract their envelopes, phase velocities are calibrated.  
These calibration runs eliminate free parameters during fitting and make the stimulated run well controlled.  
Third, competing processes are monitored to ensure that only valid data are used for fitting. Simulation parameters are chosen to reduce the impacts of spontaneous scattering, and the seed wavelength is scanned to excite leading resonances, which are mediated by the Langmuir-like P wave and the electron-cyclotron-like F wave. In addition, evolution of distribution function is monitored to select data within a time window where plasma conditions remain constant. Fitting well-controlled simulation data to analytical solutions of the same setup leads to excellent agreements in Sec.~\ref{Sec:Results}, where the warm-fluid theory is shown to be valid within a wide parameter range.
The protocol has difficulties for weak resonances, primarily due to spontaneous scattering and leakages during pump-seed separation. Potential ways to circumvent the difficulties are discussed in Sec.~\ref{Sec:Discussion}, and further investigations may find that the warm-fluid theory is valid in even wider parameter spaces. Nevertheless, kinetic effects such collisionless damping and Bernstein-like resonances are clearly observed, suggesting the importance of developing and benchmarking kinetic theories of MagLPI in the future.

%%%%%%%%%%%%%%%%%%%%%%%%%%%%%%%%%%%%%%%%%%%
%%%%%%%%%%%%%%%%%%%%%%%%%%%%%%%%%%%%%%%%%%%
\section{Analytic solutions of linearized three-wave equations}\label{Sec:Analytic}
In the slowly-varying amplitude approximation $E=\mathcal{E}\sin\theta$, where the wave envelop $\mathcal{E}$ varies slowly compared to the wave phase $\theta=\mathbf{k}\cdot\mathbf{x}-\omega t +\theta_0$, the interaction between three resonant waves, which satisfy $\omega_1=\omega_2+\omega_3$ and $\mathbf{k}_1=\mathbf{k}_2 +\mathbf{k}_3$, is described by the three-wave equations
\begin{subeqnarray}
	\label{eq:three-wave}
	d_t a_1 & = & -\frac{\Gamma}{\omega_1} a_2 a_3, \\
	d_t a_2 & = & \frac{\Gamma}{\omega_2} a_1 a_3, \\
	d_t a_3 & = & \frac{\Gamma}{\omega_3} a_1 a_2,
\end{subeqnarray}
where $d_t=\partial_t+\mathbf{v}\cdot\nabla +\mu$ is the advective derivative at group velocity $\mathbf{v}=\partial\omega/\partial\mathbf{k}$ and $\mu$ is a phenomenological damping rate. 
The above equations describe the decay of pump wave $a_1$ into daughter waves $a_2$ and $a_3$, and the inverse process. The dimensionless $a=e|\mathbfcal{E}|u^{1/2}/m_e\omega c$ is the normalized electric-field amplitude. The normalization is such that when $a>1$, the quiver velocity of electrons, whose charge is $-e$ and mass is $m_e$, becomes comparable to the speed of light $c$. The normalization also involves the wave-energy coefficient $u=\frac{1}{2}\mathbf{e}^\dagger\mathbb{H}\mathbf{e}$, where $\mathbb{H}$ is the Hamiltonian of linear waves and $\mathbf{e}$ is the unit polarization vector, such that the cycle-averaged energy of the wave is $\frac{1}{2}\epsilon_0 u\mathbfcal{E}^2$. The wave-energy coefficient is simply $u=1$ for unmagnetized electromagnetic waves and cold Langmuir waves. However, for magnetized plasma waves, $u$ usually differs from unity and can be evaluated using the code in \citet{Shi22} for given eigenmodes.

The key parameter in the three-wave equation is the coupling coefficient $\Gamma$, which has units of frequency squared. In magnetized warm-fluid plasmas, $\Gamma$ is given by the explicit formula \citep{Shi17,Shi19}
\begin{equation}
	\label{eq:coupling}
	\Gamma = \sum_s\frac{Z_s\omega_{ps}^2(\Theta^s+\Phi^s)}{4M_s(u_1u_2u_3)^{1/2}},
\end{equation}
where the summation is over all plasma species, with normalized charge $Z_s=e_s/e$, mass $M_s=m_s/m_e$, and plasma frequency $\omega_{ps}^2=e_s^2n_{s0}/\epsilon_0 m_s$ at equilibrium density $n_{s0}$. 
In the numerator, $\Theta$ is due to electromagnetic scattering and is equals to the sum of $\Theta_{1,\bar{2}\bar{3}}$ with its five permutations. Explicitly, $\Theta_{i,jl}=\frac{1}{\omega_j}(c\mathbf{k}_{i}\cdot\mathbf{f}_j)(\mathbf{e}_{i}\cdot\mathbf{f}_l)$, where $\mathbf{f}=\mathbb{F}\mathbf{e}$ and $\mathbb{F}$ is related to the linear susceptibility by $\chi=-\omega_p^2\mathbb{F}/\omega^2$, which reduces to $\mathbb{F}=\mathbb{I}$ for unmagnetized cold plasmas. The notation $\bar{i}$ in subscripts means complex conjugation for $\mathbf{e}_i$ and $\mathbf{f}_i$ with negations for the wave 4-momentum $(\omega_i, \mathbf{k}_i)$. 
Finally, the $\Phi$ term in Eq.~(\ref{eq:coupling}) is due to warm-fluid nonlinearities, which is typically smaller than $\Theta$ by a factor of $v_T^2/c^2$, where $v_T$ is thermal speed.
The coupling coefficient can be readily evaluated using the code in \citet{Shi22} once plasma conditions and the three resonant waves are specified.

To benchmark the value of magnetized three-wave coupling coefficient $\Gamma$ using numerical simulations, this paper considers a linearized and one-dimensional problem whereby the three-wave equations are simplified. First, the pump wave is launched with an amplitude that is much larger than the daughter waves, in which case $a_1$ remains approximately a constant. Second, simulations in one spatial dimension are used, meaning that $\mathbf{k}\parallel \hat{\mathbf{x}}$ and the wave envelopes only vary along the $x$ direction. Notice that the group velocity $\mathbf{v}$ can have components in other directions, but $\mathbf{v}\cdot\nabla$ only picks up its $x$ component $v$. 
With these simplifications, Eqs.~(\ref{eq:three-wave}) becomes a coupled-mode equation $\mathsfbi{L}\ubalpha=\mathbf{0}$ where 
\begin{equation}
	\label{eq:linear-three-wave}
	\mathsfbi{L} =
	\left( \begin{array}{cc}
		\partial_t+v_2\partial_x+\mu_2 & -\gamma_0 \\
		-\gamma_0 & \partial_t+v_3\partial_x+\mu_3
	\end{array} \right),
\end{equation}
%\begin{subeqnarray}
%	\label{eq:linear-three-wave}
%	(\partial_t + v_2\partial_x + \mu_2)\alpha_2 & = & \gamma_0\alpha_3, \\
%	(\partial_t + v_3\partial_x + \mu_3)\alpha_3 & = & \gamma_0\alpha_2,
%\end{subeqnarray}
and $\ubalpha=(\alpha_2, \alpha_3)^\mathrm{T}$ is a column vector. Here, $\alpha=\sqrt{\omega}a$ is rescaled such that the off diagonal elements are the same $\gamma_0$. Since only $\gamma_0^2$ is of physical significance, we can pick the positive sign so that the bare growth rate of daughter waves is
\begin{equation}
	\label{eq:growthrate}
	\gamma_0=\frac{|\Gamma a_1|}{\sqrt{\omega_2\omega_3}}. 
\end{equation}
Without loss of generality, we can always choose to label the daughter waves such that $|v_2|\ge |v_3|$. Moreover, we can always choose a coordinate such that $v_2\ge0$. With these choices, $\Delta v = v_2-v_3\ge0$ is nonnegative. Solutions to  $\mathsfbi{L}\ubalpha=\mathbf{0}$ are different in forward ($v_3>0$) and backward ($v_3<0$) scattering cases.

Since the equations are linear, they admit exponential solutions that are simple to write down analytically but difficult to set up numerically. The exponential solutions are of the form $\alpha_2\propto \alpha_3\propto \exp(\gamma t + \kappa x)$ where the temporal and spatial growth rates satisfy 
\begin{equation}
	\label{eq:dispersion}
	(\gamma+v_2\kappa+\mu_2)(\gamma+v_3\kappa+\mu_3)=\gamma_0^2. 
\end{equation} 
The above constraint defines a hyperbola in the $(\kappa, \gamma)$ plane. 
One special case is $\kappa=0$, where the wave envelopes are uniform in space. The two roots are $\gamma_\pm=-\bar{\mu}\pm\Omega$,
where $\bar{\mu}=\frac{1}{2}(\mu_2+\mu_3)$, $\Omega=\sqrt{\gamma_0^2 + (\frac{\Delta\mu}{2})^2}$, and $\Delta\mu=\mu_2-\mu_3$. 
When $\gamma_0<\gamma_c$, both roots are negative and correspond to damping. On the other hand, when $\gamma_0>\gamma_c$, one root becomes positive, giving rise to parametric instability whose threshold is
\begin{equation}
	\label{eq:parametric_threshold}
	\gamma_c = \sqrt{\mu_2\mu_3}.
\end{equation}
The other special case is $\gamma=0$, where the wave envelopes are stationary in time. Assuming $v_2v_3\ne0$, then for forward scattering $v_2v_3>0$, the envelopes decay in space unless $\gamma_0>\gamma_c$, similar to the previous case. However, the backscattering case $v_2v_3<0$ is very different: Purely growing or decaying solution no longer exists when $\gamma_0$ exceeds the absolute instability threshold
\begin{equation}
	\label{eq:absolute_threshold}
	\gamma_a = \frac{1}{2}\sqrt{|v_2v_3|}\Big(\frac{\mu_2}{|v_2|} + \frac{\mu_3}{|v_3|}\Big).
\end{equation}
When $\gamma_0>\gamma_a$, the two roots of $\kappa$ acquire imaginary parts, which means that stationary exponential solutions become oscillatory in space. The significance of $\gamma_a$ will become apparent in Eq.~(\ref{eq:alphaStep}) when we discuss the backscattering problem. 
To extract growth rates from kinetic simulations, which are usually designed to solve initial boundary value problems, one approach is to choose an initial condition that corresponds to a uniform $\alpha$ and watch it grow in time. However, the effect of growth is mixed with damping, whose rate is unknown when waves propagate at oblique angles with the magnetic field. An alternative approach is to run simulations until the system reaches steady state. However, the plasma distribution functions also evolve during the process, sometimes quite substantially \citep{Manzo22}, so the growth and damping are not only mixed but are also not constants. 
To overcome these difficulties, kinetic simulations are fitted using more general solutions of $\mathsfbi{L}\ubalpha=\mathbf{0}$ by solving initial boundary value problems, whose solutions are known \citep{bobroff1967impulse,mounaix1994space,hinkel1994temporal} but are rederived below for clarity.
The spatial variations of $\alpha$ allow the effects of growth and damping to be distinguished, and the transient-time response before plasma conditions evolve allows the growth and damping rates to be treated as constants. 

%%%%%%%%%%%%%%%%%%%%%%%%%%%%%%%%%%%%%%%%%%%
\subsection{Initial value problem}
In the initial value problem, the spatial domain is infinite, and the wave envelops evolve in time from their initial conditions
\begin{equation}
	\label{eq:initial}
	\ubalpha(x,t=0) = \mathbf{A}(x),
\end{equation}
where $\mathbf{A}=(A_2, A_3)^\mathrm{T}$.
In the degenerate case $v_2=v_3=v$, after transforming to the comoving frame $x'=x-vt$ and $t'=t$ where $\partial_t+v\partial_x=\partial_{t'}$, the equations become ordinary differential equations (ODEs) in $t'$. The eigenvalues $\gamma_\pm=-\bar{\mu}\pm\Omega$ of the linear ODEs are the $\kappa=0$ roots of Eq.~(\ref{eq:dispersion}), and the general solutions are of the form $\alpha(x',t')=A_+(x')e^{\gamma_+t'} + A_-(x')e^{\gamma_-t'}$. The coefficients $A_\pm$ are determined by matching initial conditions, which give the solution 
\begin{equation}
	\label{eq:degenerate}
	\ubalpha=e^{-\bar{\mu}t}
	\left( \begin{array}{cc}
	\cosh \Omega t-\frac{\Delta\mu}{2\Omega} \sinh \Omega t & \frac{\gamma_0}{\Omega} \sinh \Omega t \\
	\frac{\gamma_0}{\Omega} \sinh \Omega t  & \cosh \Omega t+\frac{\Delta\mu}{2\Omega} \sinh \Omega t
	\end{array} \right)
	\mathbf{A}(x-vt).
\end{equation} 
The above $\Delta v=0$ solution exhibits two features that are more general: a diagonal damping term, and off-diagonal coupling terms that vanish when $\gamma_0=0$. 
The rest of this paper will focus on the nondegenerate case $\Delta v>0$, because in LPI $v_2$ is close to the speed of light whereas $v_3$ is on the scale of thermal velocities, which are much slower.

In the nondegenerate case $\Delta v> 0$, since the equations are linear, they can be solved using Fourier transform $\mathcal{F}[p](k)=\int_{-\infty}^{+\infty} dx\; p(x)\exp(-ikx)=\hat{p}(k)$. In momentum space, the equation becomes $\partial_t\hat{\ubalpha}=\mathsfbi{K}\hat{\ubalpha}$ where 
\begin{equation}
	\label{eq:K}
	\mathsfbi{K} =
	\left( \begin{array}{cc}
		-ikv_2-\mu_2 & \gamma_0 \\
		\gamma_0 & -ikv_3 -\mu_3
	\end{array} \right).
\end{equation}
Since the matrix is time independent, the solution is $ \hat{\ubalpha}(t)=\exp(t\mathsfbi{K})\hat{\mathbf{A}}$. The matrix exponential can be computed by diagonalizing $\mathsfbi{K}$ whose eigenvalues are $\lambda_\pm=-ik\bar{v}-\bar{\mu}\pm\sqrt{\gamma_0^2-\omega^2}$, where $\bar{v}=\frac{1}{2}(v_2+v_3)$ and $\omega=\frac{1}{2}(k\Delta v-i\Delta\mu)$. Finding eigenvectors of $\lambda_\pm$ and diagonalizing $\mathsfbi{K}$, the solution map $\hat{\Phi}(k,t)=\exp(t\mathsfbi{K})$ can be written as
\begin{equation}
	\label{eq:Phik}
	\hat{\Phi}(k,t) 
	= e^{-(ik\bar{v}+\bar{\mu})t}
	\left( \begin{array}{cc}
		\partial_t-i\omega & \gamma_0 \\
		\gamma_0 & \partial_t+i\omega
	\end{array} \right)\hat{G}(k,t),
\end{equation}
where $\hat{G}(k,t) = \sinh (t\sqrt{\gamma_0^2-\omega^2})/\sqrt{\gamma_0^2-\omega^2}$.
The solution map $\hat{\Phi}(k,t)$ satisfies matrix equation $\partial_t\hat{\Phi}(k,t)=\mathsfbi{K}\hat{\Phi}(k,t)$ and initial condition $\hat{\Phi}(k,t=0)=\mathsfbi{I}$, where $\mathsfbi{I}$ is the two-dimensional identity matrix.
Notice that the behavior of $\hat{G}$ is regular when $\omega(k)\rightarrow\pm\gamma_0$. We can choose the branch cut for the square roots to lie between these two points. Since $\sinh$ is an odd function, $\hat{G}$ is in fact analytic in the complex $k$ plane.

To find the solution in $x$ space, take inverse Fourier transform $\mathcal{F}^{-1}[\hat{p}](x) =\int_{-\infty}^{+\infty} \frac{dk}{2\pi} \hat{p}(k) e^{ikx} =p(x)$. Since $\omega$ depends on $k$, it is convenient to change the integration variable to $k'=2\omega/\Delta v$, which gives $k=k'+i\Delta\mu/\Delta v$. Moreover, it is convenient to change the reference frame to
\begin{subeqnarray}
	\label{eq:COMframe}
	\tau & = & \frac{\Delta v}{2} t, \\
	z & = & x-\bar{v}t,
\end{subeqnarray} 
which travels at the average velocity of the two waves. Notice that $z-\tau=x-v_2t$ and $z+\tau=x-v_3t$. 
The phase factor simplified as $\exp[ikx-(ik\bar{v}+\bar{\mu})t]=\rho e^{ik'z}$ where 
\begin{equation}
	\label{eq:rho}
	\rho = \exp\Big\{\frac{1}{\Delta v}\Big[\mu_3(z-\tau)-\mu_2(z+\tau)\Big]\Big\}, 
\end{equation}
which is a damping factor independent of $k'$. When taking inverse Fourier transform of Eq.~(\ref{eq:Phik}), all matrix elements can be expressed in terms of 
\begin{subeqnarray}
	\label{eq:Green}
	g(z, \tau) &=& \int_{-\infty-i\epsilon}^{+\infty-i\epsilon} \frac{dk'}{2\pi}  e^{ik'z}\frac{\sinh \tau\sqrt{m^2-k'^2}}{\sqrt{m^2-k'^2}} \\
	 &=& \frac{1}{2}\mathrm{sign}(\tau)I_0\Big(m\sqrt{\tau^2-z^2}\Big)\Theta(\tau^2-z^2),
\end{subeqnarray}
where $m=2\gamma_0/\Delta v$, $I_0$ is modified Bessel function, and $\Theta$ is Heaviside step function. 
The shift by $\epsilon = \Delta\mu/\Delta v$ is insignificant because the integrand is analytic. 
The above integral is calculated in appendix~\ref{appA}. The solution map $\Phi$ in configuration space is
\begin{equation}
	\label{eq:Phix}
	\Phi(x,t) 
	= \rho
	\left( \begin{array}{cc}
		\partial_\tau-\partial_z & m \\
		m & \partial_\tau+\partial_z
	\end{array} \right)
	g(z, \tau).
\end{equation}
The derivatives are evaluated using $I'_0(\xi)=I_1(\xi)$ and $\Theta'(\xi)=\delta(\xi)$, and an explicit formula is given by Eq.~(\ref{eq:PhixExplicit}). 
The function $g$ satisfies the imaginary-mass Klein-Gordon equation $(\partial_\tau^2-\partial_z^2-m^2)g(z,\tau)=0$, as shown in Eq.~(\ref{eq:gGreen}). Consequently, the solution map satisfies $\mathsfbi{L}\Phi(x,t)=\mathsfbi{0}$ following Eq.~(\ref{eq:Drho}). 
The step function $\Theta$ enforces the causality that information outside the light cone $\tau^2-z^2=(v_2t-x)(x-v_3t)$ does not affect solutions.

Finally, to invert $\hat{\ubalpha}=\hat{\Phi}\hat{\mathbf{A}}$, compute inverse Fourier transform of products, which are given by convolutions $\mathcal{F}^{-1}[\hat{p}\hat{q}](x)=\int_{-\infty}^{+\infty} dx'\; p(x')q(x-x')$. 
Since the phase factor is simpler in $k'=k-i\Delta\mu/\Delta v$, in addition to the change of variables in Eq.~(\ref{eq:COMframe}), it is convenient to define $\hat{\mathbf{B}}(k')= \hat{\mathbf{A}}(k)$, which means that $\mathbf{B}(x) = \exp(x\Delta\mu/\Delta v)\mathbf{A}(x)$. 
With $\ubalpha=\rho\ubbeta$, where $\rho$ is given by Eq.~(\ref{eq:rho}), the solution can be written as
\begin{subeqnarray}
	\label{eq:beta}
	\beta_2(z,\tau) & = & B_2(z\!-\!\tau)+\frac{m}{2}\!\int_{-\tau}^{\tau}\!dz' \Big[B_2(z\!-\!z')\sqrt{\frac{\tau\!+\!z'}{\tau\!-\!z'}}I_1(\xi')+B_3(z\!+\!z')I_0(\xi')\Big], \quad\\
	\beta_3(z,\tau) & = & B_3(z\!+\!\tau)+\frac{m}{2}\!\int_{-\tau}^{\tau}\!dz' \Big[B_3(z\!+\!z')\sqrt{\frac{\tau\!+\!z'}{\tau\!-\!z'}}I_1(\xi')+B_2(z\!-\!z')I_0(\xi')\Big], \quad
\end{subeqnarray}
where $\xi'=m\sqrt{\tau^2-z'^2}$. An explicit expression of $\ubalpha$ in $(x,t)$ coordinate is given later in Eq.~(\ref{eq:alphaAhead}). It is straightforward to check that the above expression satisfies the differential equation and the initial conditions. When $\mathbf{B}$ only involves $\delta$ or step functions, the above integrals can be readily evaluated. However, for general initial conditions, closed-form analytical expressions may not exist, so the above integrals is evaluated numerically. Compared to numerical integration of the differential equations, which advances initial conditions step by step, the above integral solution allows fast forwarding, which directly gives the solution at desired final time. Notice that by rescaling $z'=\tau \zeta$, the numerical integration is always within the range $\zeta\in[-1,1]$, so the cost of evaluating the integral does not increase with time for smooth initial conditions.

%%%%%%%%%%%%%%%%%%%%%%%%%%%%%%%%%%%%%%%%%%%
\subsection{Initial boundary value problem}
Compared to the initial value problem, the boundary value problem is easier to set up in kinetic simulations. In the initial value problem, the distribution functions need to be specified meticulously in both the configuration space and the velocity space in order to ensure that only the desired eigenmodes are excited. In comparison, in boundary value problem, only electromagnetic fields at the boundary need to be specified. Using wave frequencies and polarizations to select excited waves, the desired eigenmodes then propagate into the initially Maxwellian plasma where interactions occur. With a single boundary at $x=0$, the initial boundary value problem is specified by the conditions
\begin{subeqnarray}
	\label{eq:boundary}
	\ubalpha(x>0,t=0) & = & \mathbf{A}(x), \\
	\ubalpha(x=0,t>0) & = & \mathbf{h}(t),
\end{subeqnarray}
where $\mathbf{h}=(h_2,h_3)^\mathrm{T}$. In the forward scattering case $v_2>v_3>0$, these conditions can be specified separately. However, in the backscattering case $v_2>0>v_3$, in order for the problem to be well-posed, only two conditions can be specified independently.

Since the equation $\mathsfbi{L}\ubalpha=\mathbf{0}$ is linear, the initial boundary value problem can be solved using Laplace transform 
$\mathcal{L}[p](k)=\int_{0}^{+\infty} dx\; p(x)\exp(-ikx)=\tilde{p}(k)$. Using property of the Laplace transform that $\tilde{p'}(k)=ik\tilde{p}(k)-p(0)$, the equation becomes $\partial_t\tilde{\ubalpha}=\mathsfbi{K}\tilde{\ubalpha}+\mathbf{H}$, where $\mathsfbi{K}$ is given by Eq.~(\ref{eq:K}) and $\mathbf{H}=(v_2h_2,v_3h_3)^\mathrm{T}$ only depends on time. 
Using the Duhamel's principle, the inhomogeneous ODE is solved by $\tilde{\ubalpha}(t)=\hat{\Phi}(t)\tilde{\mathbf{A}}+\int_0^t  dt' \hat{\Phi}(t-t')\mathbf{H}(t')$, where $\hat{\Phi}$ is given by Eq.~(\ref{eq:Phik}). 
To find the solution in configuration space, take inverse Laplace transform $\mathcal{L}^{-1}[\hat{p}\tilde{q}](x)=\int_\mathcal{C} \frac{dk}{2\pi} \mathcal{F}[p](k)\mathcal{L}[q](k) e^{ikx}=\int_0^{+\infty} dx'p(x-x')q(x')$, 
where the contour $\mathcal{C}$ runs below all poles. The solution is
\begin{equation}
	\label{eq:alpha}
	\ubalpha(x,t)=\int_0^{\infty} dx'\Phi(x-x',t)\mathbf{A}(x')+\int_0^t dt'\Phi(x,t-t')\mathbf{H}(t'),
\end{equation}
where $\Phi$ is given by Eq.~(\ref{eq:Phix}). Using $\mathsfbi{L}\Phi=\mathsfbi{0}$, the above expression clearly satisfies $\mathsfbi{L}\ubalpha=\mathbf{0}$. Moreover, using the explicit expression of $\Phi$ in Eq.~(\ref{eq:PhixExplicit}), it is easy to see that $\Phi(x,t=0)=\delta(x)\mathsfbi{I}$, so the initial conditions are always satisfied. 
%Moreover, it is easy to see that $\Phi(x=0^+,t)=\delta(t)\mathrm{diag}(1/v_2, 1/v_3)$ when $v_3>0$, so the boundary conditions are always satisfied in the forward scattering case. 
However, the situation for boundary conditions depends on the sign of $v_3$, as we shall see below.

\begin{figure}
	\centerline{\includegraphics[width=0.6\textwidth]{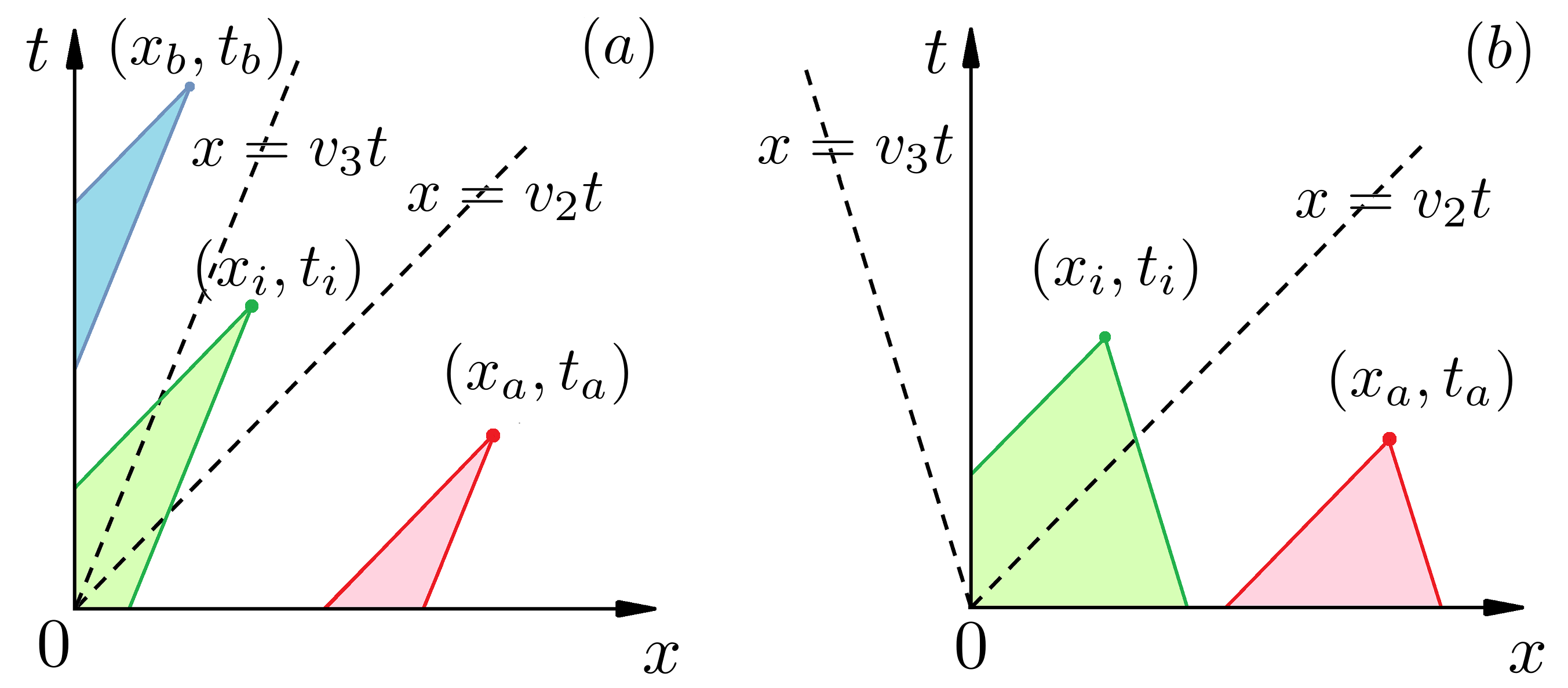}}% Images in 100% size
	\caption{Solutions are determined by initial and boundary conditions within the past light cone. For $(x_a, t_a)$ ahead of $x>v_2t$, the light cone (red) only intercepts the $x$ axis, so the solution is independent of boundary conditions. For $(x_b, t_b)$ behind $x<v_3t$, the light cone (blue) only intercepts the $t$ axis, so the solution is independent of initial conditions. For $(x_i, t_i)$ within $v_3t<x<v_2t$, the solution depends on both initial and boundary conditions. $(a)$ In the forward scattering case, initial and boundary conditions can be specified independently. $(b)$ In the backscattering case, initial conditions arrive at the boundary so constraints must be satisfied.}
	\label{fig:boundary}
\end{figure}
%$x, t, (x_a, t_a), (x_i, t_i), (x_b, t_b)$
%$x_a-v_3t_a, x_a-v_2t_a, x_i-v_2t_t$, $x=v_2t, x=v_3t$
%\begin{equation}
%	t_b-\frac{x_b}{v_2}, t_b-\frac{x_b}{v_3}, t_i-\frac{x_i}{v_2}
%\end{equation} 
%$0, (a), (b)$

Using the property that $\Phi$ is zero outside the light cone, the above integral solution, which is equivalent to $\ubalpha(x,t)=\int_{-\infty}^x dx'\Phi(x',t)\mathbf{A}(x-x')+\int_0^t dt'\Phi(x,t')\mathbf{H}(t-t')$, is simplified in the three regions shown in figure~\ref{fig:boundary}. 
First, ahead of the light cone $x>v_2t$, effects of boundary conditions have not arrived, so only the initial conditions contribute. In terms of the rescaled variable $\ubbeta$, the solution is given by Eq.~(\ref{eq:beta}), and in terms of the original variables, the solution when $x>v_2t>0$ is 
\begin{equation}
	\label{eq:alphaAhead}
	\ubalpha(x,t)=
	\left( \begin{array}{c}
		e^{-\mu_2t}A_2(x-v_2t) \\
		e^{-\mu_3t}A_3(x-v_3t)
	\end{array} \right) 
	+ 
	\frac{\gamma_0}{\Delta v}\int_{v_3t}^{v_2t} dx' \rho(x',t)\Phi_0(x',t)\mathbf{A}(x-x'),
\end{equation}
where $\rho$ is the damping factor given by Eq.~(\ref{eq:rho}) and $\Phi_0$ is the kernel within the light cone given in Eq.~(\ref{eq:PhixExplicit}). When $t=0$, the solution clearly satisfies the initial conditions. 
Second, behind the light cone $x<v_3t$, which is within the domain only when $v_3>0$, effects of the initial conditions have propagated away, so only the boundary conditions contribute. Therefore, when $v_3t>x>0$, the solution is 
\begin{equation}
	\label{eq:alphaBehind}
	\ubalpha(x,t)=
	\left( \begin{array}{c}
		e^{-\mu_2x/v_2}h_2(t-x/v_2) \\
		e^{-\mu_3x/v_3}h_3(t-x/v_3)
	\end{array} \right) 
	+
	\frac{\gamma_0}{\Delta v}\int_{x/v_2}^{x/v_3} dt' \rho(x,t') \Phi_0(x,t') \mathbf{H}(t-t').
\end{equation}
The solution clearly satisfies the boundary conditions at $x=0$ for this forward scattering case. 
Finally, inside the light cone, both the initial and boundary conditions contribute, and the solution when  $v_3t<x<v_2t$ is given by 
\begin{eqnarray}
	\label{eq:alphaWithin}
	\nonumber
	\ubalpha(x,t) &=&
	\left( \begin{array}{c}
		e^{-\mu_2x/v_2}h_2(t-x/v_2) \\
		e^{-\mu_3t}A_3(x-v_3t)
	\end{array} \right)  \\
	&+& \frac{\gamma_0}{\Delta v}\int_{v_3t}^{x} dx' \rho(x',t) \Phi_0(x',t) \mathbf{A}(x-x') \\
	\nonumber
	&+& \frac{\gamma_0}{\Delta v}\int_{x/v_2}^{t} dt' \rho(x,t') \Phi_0(x,t') \mathbf{H}(t-t').
\end{eqnarray}
In the forward scattering case $v_3>0$, the future light cone is within the domain, so the initial and boundary conditions can be specified independently. However, in the backscattering case $v_3<0$, the future light cone is intercepted by the boundary. Intuitively, when information propagates towards left and arrives at the boundary, one cannot arbitrarily set values at the boundary.

%%%%%%%%%%%%%%%%%%%%%%%%%%%%%%%%%%%%%%%%%%%
\subsection{Backscattering problem\label{sec:BS}}
When the initial conditions are zero, as will be the case in kinetic simulations, the integral constraints $\ubalpha(x=0,t>0)=\mathbf{h}(t)$ for the backscattering case $v_3<0$ can be solved explicitly. By setting $x=0$ and $\mathbf{A}=\mathbf{0}$ in Eq.~(\ref{eq:alphaWithin}), the constraints can be simplified as $(0, l_3(s))^\mathrm{T}=\int_0^s ds'\Psi_0(s')\mathbf{l}(s-s')$, where $s=\gamma t$ is time normalized by $\gamma=2\gamma_0\sqrt{v_2|v_3|}/\Delta v$, $\mathbf{l}(s)=e^{\mu t}h(t)$ is rescaled by damping $\mu=(\mu_3v_2+\mu_2|v_3|)/\Delta v$, and
\begin{equation}
	\label{eq:Psis}
	\Psi_0(s)
	= \frac{1}{2}
	\left( \begin{array}{cc}
		I_1(s) & -\sqrt{\frac{|v_3|}{v_2}} I_0(s) \\
		\sqrt{\frac{v_2}{|v_3|}} I_0(s) & -I_1(s)
	\end{array} \right).
\end{equation}
Notice that $\mu t=s\gamma_a/\gamma_0$, where $\gamma_a$ is the absolute instability threshold given by Eq.~(\ref{eq:absolute_threshold}).
Since the constraint is a convolution, it becomes a product after taking Laplace transform, which gives $(0, \tilde{l}_3(\omega))^\mathrm{T} =\tilde{\Psi}_0(\omega)\tilde{\mathbf{l}}(\omega)$. 
To compute $\tilde{\Psi}_0$, use integral representation of modified Bessel function \citep[Eq.~10.32.2]{DLMF} that $I_0(s)=\frac{1}{\pi}\int_{-1}^1 dt \, e^{-st}/\sqrt{1-t^2}$ and perform the $s$ integral first, which gives $\tilde{I}_0(\sigma) = 1/\sqrt{\sigma^2-1}$ where $\sigma=i\omega$. Then, $\tilde{I}_1(\sigma) = \sigma/\sqrt{\sigma^2-1}-1$ because $I_1(s)=I'_0(s)$.  
Since $I_n(s)\simeq e^s/\sqrt{2\pi s}$ when $s\rightarrow\infty$, the Laplace transforms converge only when Re$(\sigma)>1$. 
Solving the constraint in frequency domain gives a unique solution 
\begin{equation}
	\label{eq:l3w}
	\tilde{l}_3(\omega) = \sqrt{\frac{v_2}{|v_3|}}\Big(\sigma-\sqrt{\sigma^2-1}\Big) \tilde{l}_2(\omega),
\end{equation}
which means that if we specify the boundary condition for $\alpha_2$, then the boundary condition for $\alpha_3$ is completely determined. Taking inverse Laplace transform, whose details are shown in appendix~\ref{appB}, the self-consistent boundary condition is
\begin{equation}
	\label{eq:l3s}
	l_3(s)= \sqrt{\frac{v_2}{|v_3|}}\int_0^s ds'\frac{I_1(s')}{s'} l_2(s-s').
\end{equation}
When the normalized time $s=0$, the boundary condition $l_3(0)=0$ is initially quiescent. At later time, $\alpha_3$ builds up due to wave growth and advection. 
As shown in appendix~\ref{appB}, we can also express $l_2$ in terms of $l_3$. However, in kinetic simulations, it is much easier to specify the boundary conditions for electromagnetic waves and let the plasma evolve to fulfill the above boundary condition.

\begin{figure}
	\centerline{\includegraphics[width=0.9\textwidth]{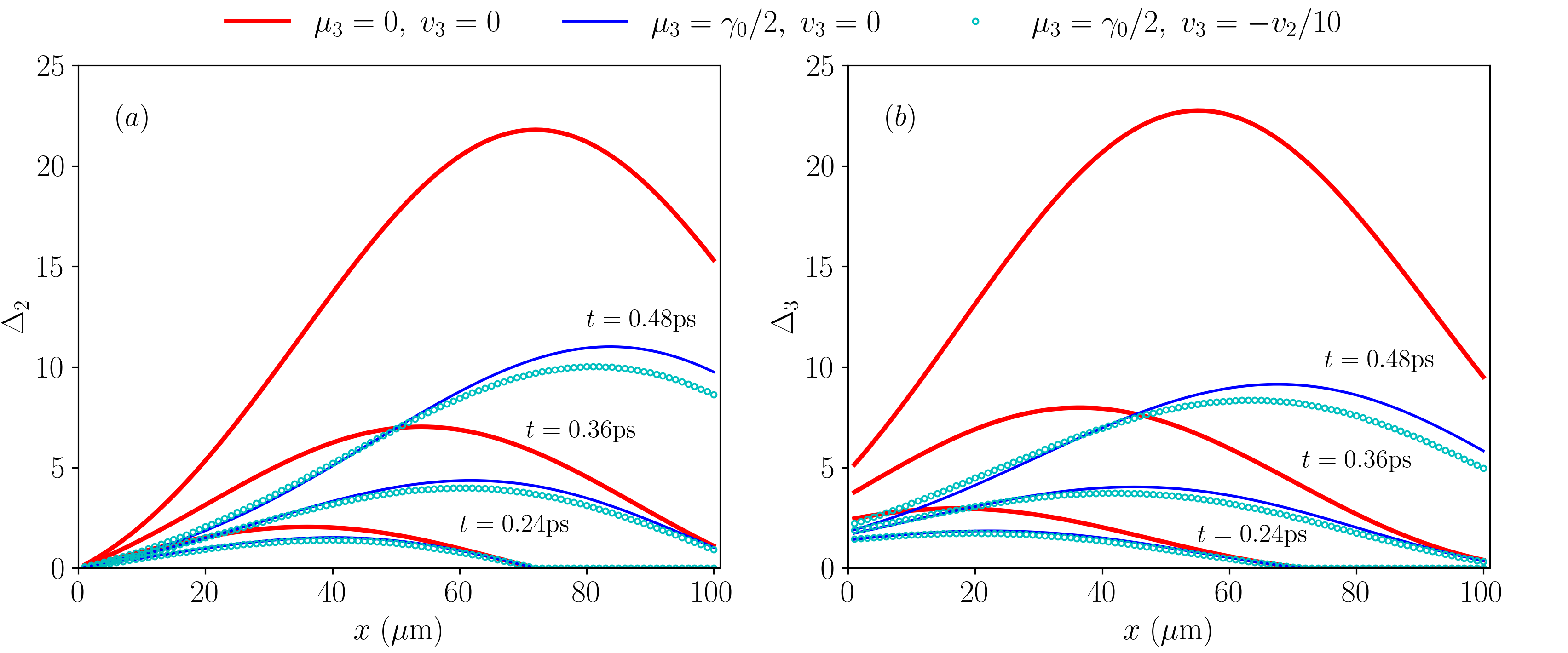}}% Images in 100% size
	\caption{The growth function $\Delta_j(x)$ in the step-function backscattering problem for $j=2$ $(a)$ and $j=3$ $(b)$ at selected time slices when \mbox{$v_2=3\times10^8$ m/s}, \mbox{$\gamma_0=10^{13}$ rad/s}, and $\mu_2=0$. 
	As time increases, $\Delta_j$ propagates in space and grows in amplitude. The growth is always zero ahead of the wave front $x=v_2t$. Moreover, $\Delta_2(0)$ is always zero due to the boundary condition. In contrast, $\Delta_3(0)$ builds up from zero as time increases. 
	Compared to the dampingless case (red lines), having an appreciable $\mu_3$ (blue lines) reduces the growth. Compared to the $v_3=0$ case (blue lines), having a small $v_3$ (cyan circles) slightly affects the solutions at early time. At later time, the discrepancies build up because $\gamma_0\approx 1.3\gamma_a$ exceeds the absolute instability threshold. 
	}
	\label{fig:Delta_spatial}
\end{figure}

\begin{figure}
	\centerline{\includegraphics[width=0.9\textwidth]{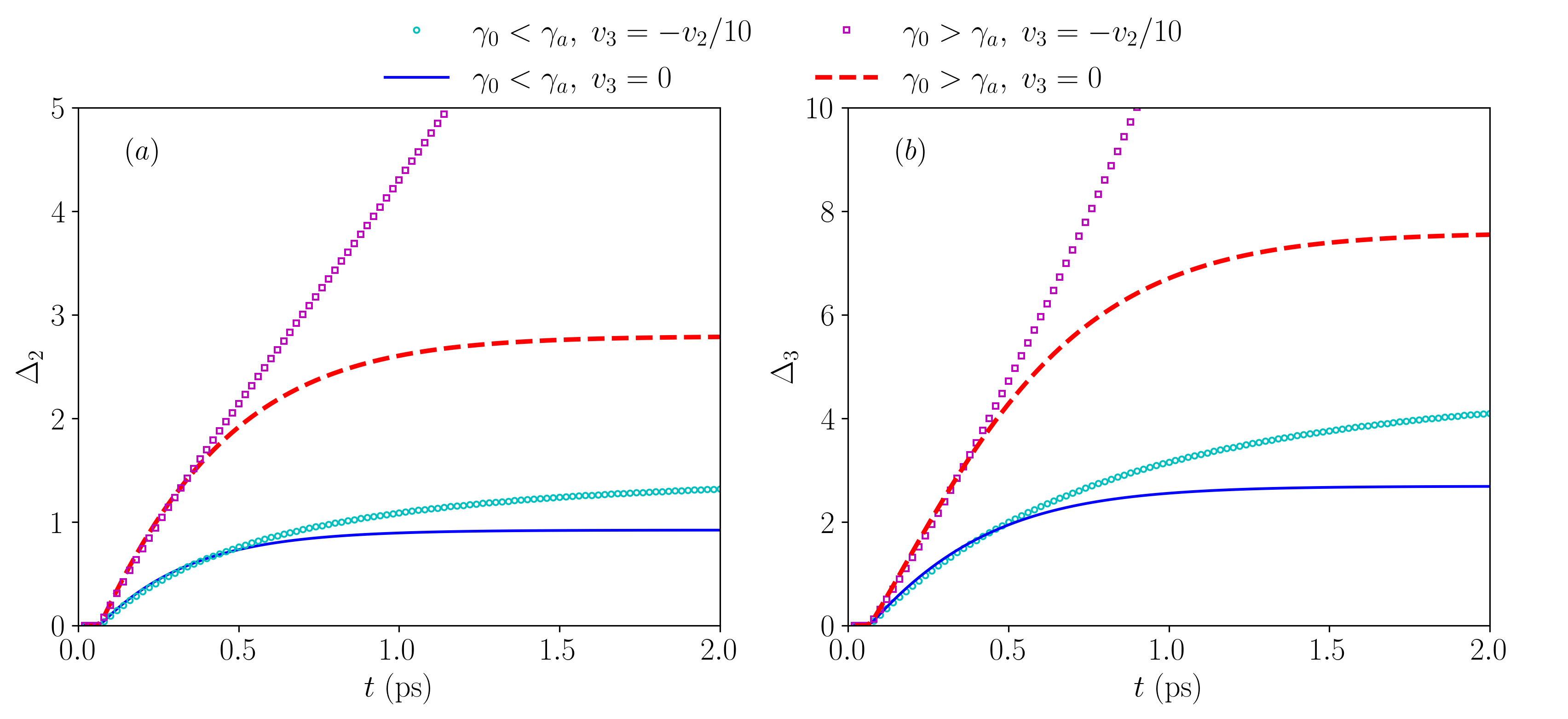}}% Images in 100% size
	\caption{The growth function $\Delta_j(t)$ in the step-function backscattering growth problem for $j=2$ $(a)$ and $j=3$ $(b)$ at $x=20\;\mu$m when \mbox{$v_2=3\times10^8$ m/s}, \mbox{$\mu_3=5\times10^{12}$ rad/s}, and $\mu_2=0$. 
	When $v_3=-v_2/10$ (symbols), the absolute instability threshold is \mbox{$\gamma_a\approx7.8\times10^{12}$ rad/s}. When \mbox{$\gamma_0=7\times10^{12}$ rad/s} (cyan) is below the threshold, the growth approaches steady state. In contrary, when \mbox{$\gamma_0=10^{13}$ rad/s} (magenta) exceeds the threshold, $\Delta_j$ continues to increase.
	When $v_3=0$ (lines), $\gamma_a$ becomes infinite so the growth always saturates. The case with \mbox{$\gamma_0=10^{13}$ rad/s} (red) has a larger steady-state value than the case with \mbox{$\gamma_0=7\times10^{12}$ rad/s} (blue).
	The effects of a small but finite $v_3$ only become significant at later time. 
	}
	\label{fig:Delta_temporal}
\end{figure}

Having expressed $h_3$ in terms of $h_2$, the solution of $\ubalpha$ is a functional of $h_2$ only. The integrals simplify when $h_2$ is $\delta$ or $\Theta$ functions. Since $\delta$ functions cannot be resolved numerically, let us focus on the case $h_2(t)=h_0\Theta(t)$, which will be used later to set up kinetic simulations. 
Using Eq.~(\ref{eq:l3s}) with $l_2(s)=h_0e^{s\gamma_a/\gamma_0} \Theta(s)$, $h_3(t)=h_0\sqrt{v_2/|v_3|}\int_0^{\gamma t} ds' e^{-s'\gamma_a/\gamma_0} I_1(s')/s'$. 
Substituting $\mathbf{h}$ into Eq.~(\ref{eq:alphaWithin}), changing integration variable to $\varphi'=\gamma(t'-x/v_2)$, and rotating the triangular double integral is a way that leads to Eq.~(\ref{eq:IdentityInverse}) gives 
\begin{subeqnarray}
	\label{eq:alphaStep}
	\alpha_2(x<v_2t) &=& h_0e^{-\mu_2x/v_2}\Big[1+\frac{1}{2}\int_0^{\gamma(t-x/v_2)} \!d\varphi\; e^{-\varphi\gamma_a/\gamma_0} D_2(\varphi, \vartheta)\Big],\\
	\alpha_3(x<v_2t) &=& h_0e^{-\mu_2x/v_2}\sqrt{\frac{v_2}{|v_3|}}\frac{1}{2}\int_0^{\gamma(t-x/v_2)} \!d\varphi\; e^{-\varphi\gamma_a/\gamma_0} D_3(\varphi, \vartheta),
\end{subeqnarray}
whereas $\ubalpha(x>v_2t)=\mathbf{0}$ ahead of the wave front.
In the above expressions, $\gamma=2\gamma_0\sqrt{v_2|v_3|}/\Delta v$, $\vartheta=\gamma_0x/\sqrt{v_2|v_3|}$, $D_2(\varphi, \vartheta) = \sqrt{1+2\vartheta/\varphi}\,I_1(\sqrt{\varphi^2+2\varphi\vartheta})-M_2(\varphi, \vartheta)$, and $D_3(\varphi, \vartheta) = I_0(\sqrt{\varphi^2+2\varphi\vartheta})-M_3(\varphi, \vartheta)$, where the kernel functions are
\begin{subeqnarray}
	\label{eq:Mfunction}
	M_2(\varphi, \vartheta) &=& \int_0^1 dr\; I_0(\sqrt{r^2\varphi^2+2r\varphi\vartheta}) \frac{I_1(\varphi(1-r))}{1-r},\\
	M_3(\varphi, \vartheta) &=& \int_0^1 dr\; \frac{I_1(\sqrt{r^2\varphi^2+2r\varphi\vartheta})}{\sqrt{1+2\vartheta/r\varphi}} \frac{I_1(\varphi(1-r))}{1-r}.	
\end{subeqnarray}
The differential properties of the kernel functions [Eqs.~(\ref{eq:Derivative2}) and (\ref{eq:Derivative3})] ensures that Eq.~(\ref{eq:alphaStep}) satisfies $\mathsfbi{L}\ubalpha=\mathbf{0}$.
Moreover, the special values $D_2(\varphi, 0)=0$ and $D_3(\varphi, 0)=2I_1(\varphi)/\varphi$ [Eqs.~(\ref{eq:M2}) and (\ref{eq:M3})] ensure that the boundary conditions are satisfied.
Using the special value, $\alpha_3\rightarrow+\infty$ when $t\rightarrow+\infty$ at $x=0$ if $\varsigma=\gamma_a/\gamma_0<1$, whereas $\alpha_3\rightarrow h_0\sqrt{v_2/|v_3|}(\varsigma - \sqrt{\varsigma^2-1})$ if $\varsigma\ge1$.
More generally when $x\ge0$, the solutions approach steady state if and only if $\gamma_0$ does not exceed the absolute instability threshold. 
The above solutions are of the form $\alpha_2=h_0 e^{-\mu_2x/v_2}(1+\Delta_2)$ and $\alpha_3=h_0e^{-\mu_2x/v_2}\Delta_3$. 
Examples of the growth function $\Delta_j$ are plotted in figures~\ref{fig:Delta_spatial} and \ref{fig:Delta_temporal}.

The integral solutions for the step-function problem are greatly simplified when $v_3\rightarrow 0$, which is a good approximation for LPI where $v_2\gg|v_3|$. 
Inside the domain, $\vartheta\rightarrow\infty$ but $\gamma\rightarrow 0$, while $\gamma\vartheta=2\gamma_0^2x/v_2$ is finite.
Since the integrands in Eq.~(\ref{eq:Mfunction}) approach zero when $\varphi\rightarrow0$ at fixed $\varphi\vartheta$, both $M_2$ and $M_3$ become zero. Then, $D_2\rightarrow \xi I_1(\xi)/\varphi$ and $D_3\rightarrow I_0(\xi)$, where $\xi=\sqrt{2\varphi\vartheta}$.
Writing integrals in Eq.~(\ref{eq:alphaStep}) in terms of $\xi$, 
\begin{subeqnarray}
	\label{eq:DeltaStep}
	\Delta_2 &=& \int_0^\psi d\xi\; e^{-\nu\xi^2}I_1(\xi) , \\
	\Delta_3 &=& \frac{v_2}{2\gamma_0 x}\int_0^\psi d\xi\; \xi e^{-\nu\xi^2}I_0(\xi),
\end{subeqnarray}
where $\psi=2\gamma_0\sqrt{(t-x/v_2)x/v_2}$ and $\nu=\mu_3v_2/4x\gamma_0^2$. Observe that $\psi=\sqrt{\mu_3t_r/\nu}$ where $t_r=t-x/v_2$ is the retarded time since the wave front passes, and $1/4\nu$ is the spatial gain exponent in steady state when $\mu_2=0$. 
After integration by part, $\Delta_3=(\gamma_0/\mu_3)[1-I_0(\psi) e^{-\nu\psi^2}+\Delta_2]$.
It is straightforward to verify that Eq.~(\ref{eq:DeltaStep}) solves $\mathsfbi{L}\ubalpha=\mathbf{0}$ when $v_3=0$. 
When $t\rightarrow+\infty$, the solutions approach steady states, which are always finite because $\gamma_a\rightarrow\infty$. Using Gaussian integrals of modified Bessel function \citep[Eqs.~10.43.24]{DLMF}, $\Delta_2=\exp(1/4\nu)-1-\mathcal{R}$. As shown in appendix~\ref{appC}, the residual $\mathcal{R}=\int_\psi^{+\infty}d\xi\, e^{-\nu\xi^2}I_1(\xi)$ decays as $e^{-\mu_3t_r}$ when $\mu_3t_r\gg\max(1,\gamma_0^2x/\mu_3v_2)$. 
The steady states $\alpha_2(x,+\infty)=(\mu_3/\gamma_0)\alpha_3(x,+\infty)=h_0 e^{\kappa x}$, where $\kappa=(\gamma_0^2/\mu_3-\mu_2)/v_2$ is consistent with linear stability analysis of Eq.~(\ref{eq:dispersion}).
The formulas in Eq.~(\ref{eq:DeltaStep}) are further simplified in two limiting cases. 
(1) When $\nu\rightarrow+\infty$, namely, when spatial gain is negligible, integrals are dominated by values near $\xi\simeq0$. The growths $\Delta_2\simeq(\gamma_0^2x/\mu_3v_2)(1-e^{-\mu_3t_r})\rightarrow0$ and $\Delta_3\simeq (\gamma_0/\mu_3)(1-e^{-\mu_3t_r})$. 
(2) When $\mu_3\rightarrow 0$, the Gaussian weight becomes unity. Using properties of modified Bessel function \citep[Eq.~10.43.1]{DLMF}, the integrals are evaluated to $\Delta_2=I_0(\psi)-1$ and $\Delta_3=(v_2/2\gamma_0 x)\psi I_1(\psi)$. 
When $\psi\simeq0$, which occurs near the boundary or the wave front, $\Delta_2\simeq \gamma_0^2t_rx/v_2$ and $\Delta_3\simeq\gamma_0 t_r$ grow linearly in time.
At given $t$, the maximum of $1+\Delta_2$ is attained at $x=\frac{1}{2}v_2 t$, which propagates at half the wave group velocity. The maximum value attained at $\psi=\gamma_0t$ is $I_0(\gamma_0 t)\simeq e^{\gamma_0t}/\sqrt{2\pi\gamma_0t}$ when $t\rightarrow+\infty$. While the exponential growth $e^{\gamma_0t}$ is intuitive, the suppression by $1/\sqrt{2\pi\gamma_0t}$ is perhaps not one would naively expect from linear instability analysis.

%%%%%%%%%%%%%%%%%%%%%%%%%%%%%%%%%%%%%%%%%%%
%%%%%%%%%%%%%%%%%%%%%%%%%%%%%%%%%%%%%%%%%%%
\section{Kinetic simulations of stimulated backscattering}\label{Sec:Simulations}
To benchmark the formula for magnetized three-wave coupling coefficient in the backscattering geometry, analytic solutions of the step-function problem are used to fit kinetic simulations in the same setup, where the simulations are performed using the PIC code EPOCH \citep{arber2015contemporary}. 
For the step-function problem, the initial condition is simply a quiescent Maxwellian plasma, whose density is chosen to be $n_e=n_i=n_0=$\mbox{$10^{19}\;\mathrm{cm}^{-3}$} and temperature $T_e=T_i=T_0$ will be scanned. The two species have the mass ratio $M_i=m_i/m_e=1837$ of hydrogen plasmas.
In the one dimensional simulation domain, the plasma occupies $x\in[0,L_p]$ with a constant $n_0$ and $T_0$, and two vacuum gaps each of length $L_v$ are placed on either side, where $L_p=80\lambda_1$, $L_v=10\lambda_1$, and \mbox{$\lambda_1=1\,\mu$m} is the vacuum pump wavelength. The slowly-varying envelope approximation requires that $L_p\gg \lambda_1$.
A constant magnetic field of strength $B_0$ is applied in the $x$-$z$ plane at an angle $\theta_B$ with respect to the $x$ axis. The special case $B_0=0$ is unmagnetized, and the special angle $\theta_B=0$ means wave vectors are parallel to the magnetic field. Both $B_0$ and $\theta_B$ will be scanned.

To achieve a constant pump amplitude, the laser is launched from the right domain boundary and ramped up from zero using a $\tanh$ profile whose temporal width equals to the laser period. The smooth ramp reduces oscillations due to numerical artifacts yet is fast enough to be viewed as a step function for the slowly varying envelope.
After propagating across the vacuum gap, most pump energy transmits into the plasma, and a small fraction is reflected from the plasma-vacuum boundary. The reflected pump leaves the domain from its right boundary, and the transmitted pump amplitude is measured from simulation data. Using analytical wave energy coefficient, the pump amplitude is normalized to $a_1$, which enters the growth rate $\gamma_0$ in Eq.~(\ref{eq:growthrate}). 
When the pump reaches the plasma-vacuum boundary on the left, most energy exists the plasma, but a small fraction is reflected. Since its wavevector is flipped, the reflected pump does not interact with the seed laser resonantly but co-propagates with the seed in $+x$ direction.

\begin{figure}
	\centerline{\includegraphics[width=1.0\textwidth]{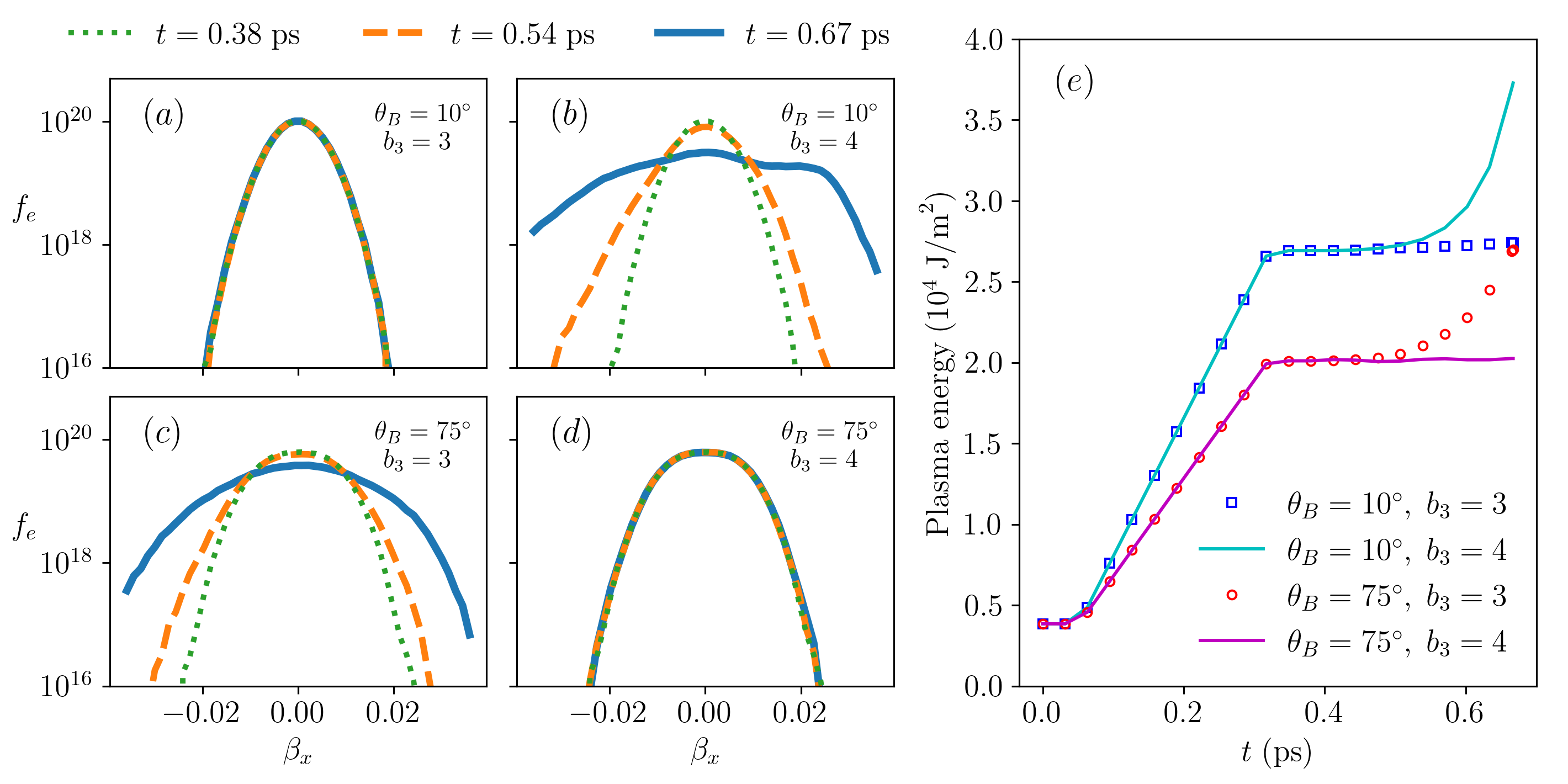}}% Images in 100% size
	\caption{Plasma evolves as pump fills and seed grows. 
	$(a)$ When $\theta_B=10^\circ$ and the coupling is weak, distribution function $f_e(\beta_x)$, where $\beta_x=v_x/c$, stays near the initial Maxwellian. 
	$(b)$ When $\theta_B=10^\circ$ but the coupling is strong, $f_e$ broadens rapidly. At \mbox{$t=0.38$ ps} (dotted green), $f_e$ is still close to Maxwellian, but when \mbox{$t=0.54$ ps} (dashed orange), non-Maxwellian tails develop. By \mbox{$t=0.67$ ps} (solid blue), a plateau-like structure resembles Landau damping in unmagnetized plasmas.
	$(c)$ When $\theta_B=75^\circ$ and the coupling is strong, $f_e$ also broadens significantly. In this case, $f_e$ remains largely symmetric due to mixing effects of gyro motion. 
	$(d)$ When $\theta_B=75^\circ$ but the coupling is weak, $f_e$ barely evolves but is no longer Maxwellian due to longitudinal quiver motion in the pump laser. 	
	$(e)$ The plasma areal energy density is initially thermal $U_T=3T_0n_0L_p$\mbox{$\approx0.38\times 10^4\; \mathrm{J/m}^2$}. As the pump fills in, plasma energy increases due to quiver motion. In unmagnetized plasma,  $U_Q=\frac{1}{2}m_ec^2a_1^2n_0L_p$\mbox{$\approx2.4\times 10^4\; \mathrm{J/m}^2$}. However, due to magnetization, the increase depends on $\theta_B$. After the seed enters, energy increases further if plasma waves are strongly excited. 
	In these examples, \mbox{$B_0=3$ kT}, \mbox{$T_0=10$ eV}, \mbox{$I_1=10^{15}\;\mathrm{W/cm}^2$}, and \mbox{$I_2=5\times10^{12}\;\mathrm{W/cm}^2$}. In $(a)$ and $(c)$, the interactions are mediated by F waves whose branch index is $b_3=3$, namely, the third highest frequency branch in warm-fluid dispersion relation. In $(b)$ and $(d)$, the interactions are mediated by P waves with $b_3=4$. 
	}
	\label{fig:SelectionEnergy}
\end{figure}

The seed laser $\alpha_2$ is launched from the left domain boundary using a similar profile but with a time delay such that its wave front enters the plasma after the pump front has existed. To measure the transmitted seed amplitude into the plasma, a seed-only run is performed to fit the boundary condition $h_0$ in the step-function problem. After the calibration steps, a stimulated run is performed where both the pump and seed lasers are turned on. 
The simulation is terminated slightly before the seed front reaches $x=L_p$, so that the analytical solution, which is obtained when only the left boundary is present, remains applicable. In addition, both the electron distribution function $f_e(v_x)$ and the pump wave amplitude $a_1(x)$ are monitored during the simulations. In many cases, the change of $f_e$ and $a_1$ are small.
For example, when the three-wave coupling is weak and the propagation angle $\theta_B$ is small (figure~\ref{fig:SelectionEnergy}$a$), the distribution function stays close to the initial Maxwellian. 
In comparison, when the coupling is week but angle is close to $90^\circ$ (figure~\ref{fig:SelectionEnergy}$d$), even though $f_e$ remains largely constant during the interaction, it is broadened from the Maxwellian due to quiver motion in the pump laser, which has an appreciable longitudinal component. 
The situation is different when the coupling is strong. Regardless of the angle, when a large amplitude plasma wave is excited, its collisionless damping leads to substantial broadening of $f_e$ as shown in figure~\ref{fig:SelectionEnergy}$(b)$ and \ref{fig:SelectionEnergy}$(c)$, which correlates with a significant increase of the plasma energy (figure~\ref{fig:SelectionEnergy}$e$) and a rapid depletion of the pump laser.
The simulation data after the peak of $f_e$ reduces by $5\%$ or $a_1$ drop by $1\%$ since interactions begin are excluded from fitting, which assumes constant plasma conditions and pump amplitude.

The choice of pump and seed laser intensities are constrained by three factors. 
First, numerical noise of the PIC method gives an upper bound of the pump intensity due to spontaneous scattering. With a finite number of sampling particles, the plasma density fluctuates around $n_0$, leading to a noise $\delta E_\parallel$ that can spontaneously scatter the pump laser $a_1$. For scatterings mediated by electron modes, the growth rate is typically comparable to the cold Raman backscattering rate $\gamma_R=\frac{1}{2}a_1(\omega_p\omega_1)^{1/2}$. The requirement $\gamma_R t_p\lesssim 1$ that noise does not grow substantially gives an upper bound for the pump intensity, where $t_p\simeq L_p/c$ is the time it takes for the pump to fill the plasma. Since $\gamma_R t_p=\pi(n_0/n_c)^{1/4}(L_p/\lambda_1) a_1$, where $n_c=\epsilon_0m_e\omega^2/e^2$ \mbox{$\approx1.1\times 10^{21}\lambda_{\mu\mathrm{m}}^{-2}\;\mathrm{cm}^{-3}$} is the critical density, we need $a_1\ll10^{-1}$ with earlier choices of $n_0$ and $L_p$. In terms of laser intensity, since \mbox{$a_1\simeq8.6\times 10^{-3} \lambda_{\mu\mathrm{m}} I_{14}^{1/2}$}, where $I_{14}$ is the intensity in units of \mbox{$10^{14}\;\mathrm{W/cm}^2$}, we see that the pump intensity cannot far exceed $I_{14}\sim10^2$.
Second, PIC noise also imposes a lower bound for the seed intensity. In each cell, the distribution function is sampled with $N$ super particles. Even though the mean velocity is zero, the standard error of the mean is $\delta v=v_T/\sqrt{N}$, where \mbox{$v_T/c\approx6.3\times10^{-2}T_{\mathrm{keV}}^{1/2}$} is the electron thermal speed. The sampling error leads to a noise current density $\delta j=en_0\delta v$, which drives a noise field $\delta E_\perp=\delta j/\epsilon_0\omega$ at frequency $\omega$ through the Amp\`{e}re's law. In terms of the relativistic critical field $E_c=m_e\omega c/e\approx$ \mbox{$3.2\times10^{12}\lambda_{\mu\mathrm{m}}^{-1}\;\mathrm{V/m}$}, the error field at the laser frequency is $\delta E_\perp=E_c (n_0 v_T/n_c c)N^{-1/2}$\mbox{$\sim10^7$ V/m} when \mbox{$T_0=10$ eV} and $N=10^2$, as shown in the inset of figure~\ref{fig:LinearEigen}$(c)$. Since the laser electric field is $E=(2I/\epsilon_0c)^{1/2}\approx$ \mbox{$2.7\times 10^{10} I_{14}^{1/2}$ V/m}, the condition $E\gg\delta E_\perp$ requires that the seed intensity be larger than $I_{14}\sim10^{-4}$, especially when the plasma is hotter.
Finally, the pump and seed amplitudes need to be separated by about an order of magnitude. This is because in order for $a_1$ to remain largely constant during three-wave interactions, we need $a_1\gg a_2$. However, if $a_1/a_2$ is too large, then filtering out $a_2$ from simulation data becomes challenging due to leakages of $a_1$ through numerical filters. Combining the three constraints, intensities of pump $I_1$ and seed $I_2$ should satisfy \mbox{$10^{10}\;\mathrm{W/cm}^2$} $\ll I_2\sim10^2 I_1$ and $I_1\ll$ \mbox{$10^{16}\;\mathrm{W/cm}^2$}. The bounds can be extended using a larger $N$. However, the benefit only increases as $\sqrt{N}$ but the numerical cost grows with $N$ linearly.

With the simulation setup and general considerations discussed above, technical details are elaborated below with examples, and the data analysis protocol is summarized in \citet{Shi22linear}.
All reported results use the resolution of $40$ cells per pump laser wavelength and $N=100$ particles per cell. Increasing or decreasing these parameters by a factor of two does not significantly change results. A larger plasma and simulation box increases the vulnerability to spontaneous pump scattering, which first shows up as unwanted oscillations on the right plasma boundary. On the other hand, using a much smaller plasma starts to violate $L_p\gg \lambda_1$, which is required in order for the slowly varying amplitude approximation to be valid.

%%%%%%%%%%%%%%%%%%%%%%%%%%%%%%%%%%%%%%%%%%%
\subsection{Launching linear eigenmodes}
Since three-wave equations describe amplitudes of linear eigenmodes, the lasers need to be launched with specified polarizations to excite targeted eigenmodes only. While polarization matching is trivial for unmagnetized plasmas, where the two electromagnetic eigenmodes are degenerate, special care needs to be taken in magnetized cases where the R and L elliptically polarized eigenmodes are nondegenerate. If the polarization is not matched properly, both R and L waves will be excited, giving rise to four polarization combinations with very different couplings \citep{shi2019amplification}.
Since incident lasers are purely transverse in the vacuum region, only the perpendicular components need to be matched with plasma eigenmodes. Denoting $\psi$ the elliptical polarization angle and $\theta$ is the wave phase, 
\begin{equation}
	\label{eq:Eperp}
	\mathbf{E}_\perp\propto \hat{\mathbf{y}}\sin\psi\sin\theta + \hat{\mathbf{k}}\times\hat{\mathbf{y}}\cos\psi\cos\theta,
\end{equation}
where $\hat{\mathbf{y}}$ and $\hat{\mathbf{k}}$ are unit vectors. The expression is symmetric about the $x$-$z$ plane where $\mathbf{B}_0$ lies. 
The special value $\psi=0$ means that the wave is linearly polarized along $\hat{\mathbf{k}}\times\hat{\mathbf{y}}$, whereas $\psi=\pi/2$ means linear polarization along $\hat{\mathbf{y}}$, $\psi=\pi/4$ is R circular polarization about $\hat{\mathbf{k}}$, and $\psi=-\pi/4$ is L circular polarization.

\begin{figure}
	\centerline{\includegraphics[width=1.0\textwidth]{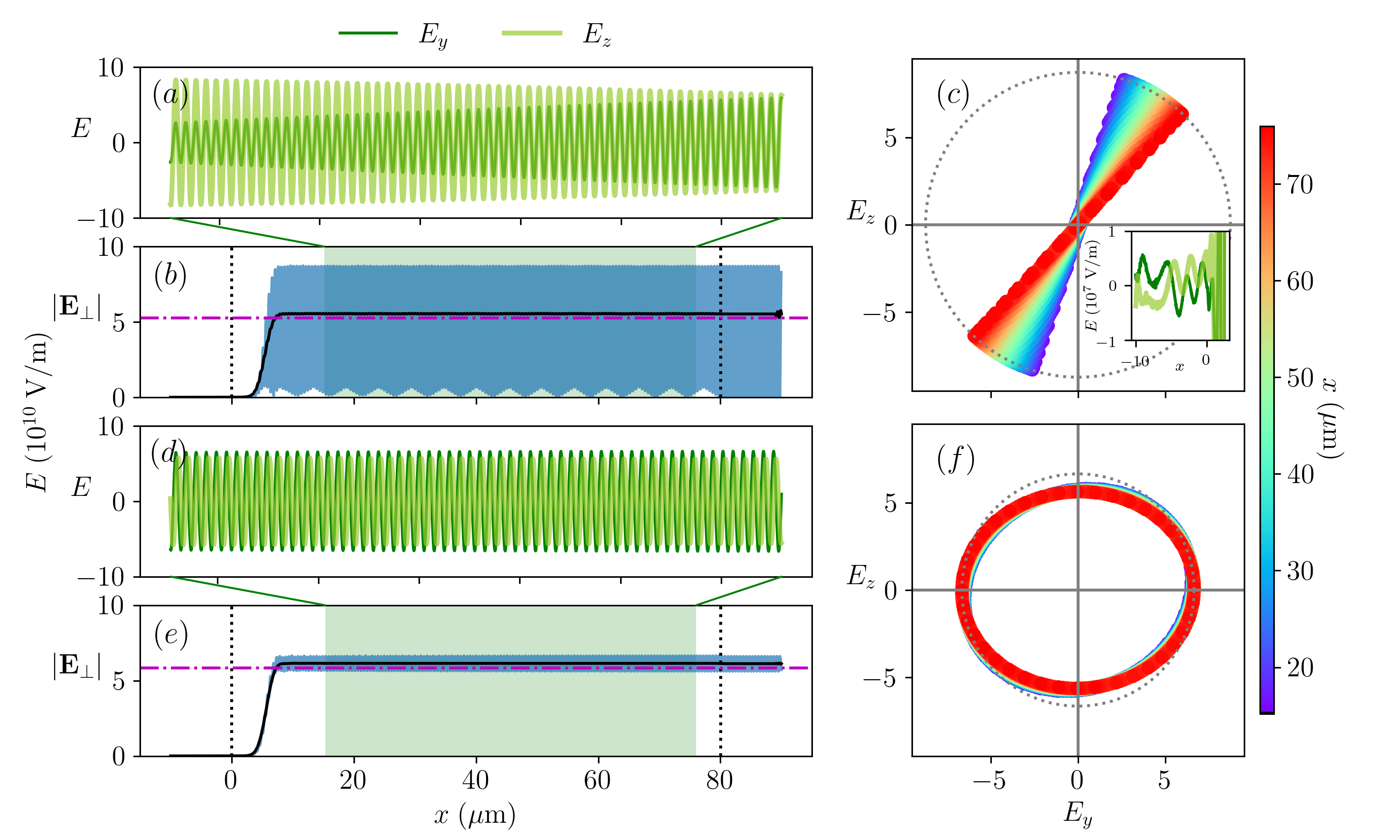}}% Images in 100% size
	\caption{By launching a linearly polarized pump laser and fitting its $E_y$ (dark green) and $E_z$ (light green) components $(a)$ within the data region $(b)$, polarization angles of numerical eigenmodes can be determined. While the polarization vector slowly rotates for none eigenmodes $(c)$ as the wave propagates in $x$ (color scale), it stays close a constant ellipse $(f)$ when an eigenmode is launched, whose field components have constant amplitude $(d)$ within the data region $(e)$. 
	The vertical dashed black lines in $(b)$ and $(e)$ mark the initial plasma-vacuum boundaries. The green shaded region is selected based on criteria that it is $4\lambda$ within the boundaries, $8\lambda$ behind the wave front, and that the $2\lambda$-running average of $|\mathbf{E}_\perp|$ (black) ramps above $95\%$ of its average (magenta). 
	The inset in $(c)$ shows the noise level before the pump arrives, which is consistent with expected PIC noise when \mbox{$T_0=10$ eV}. The polarization rotation is consistent with \mbox{$B_0=3$ kT} and $\theta_B=30^\circ$. The pump intensity is \mbox{$I_1=10^{15}\;\mathrm{W/cm}^2$}.
	}
	\label{fig:LinearEigen}
\end{figure}

Due to numerical dispersion and PIC density fluctuations, the polarization angles of numerical eigenmodes are slightly different from analytic results. Launching waves with analytical $\psi$ leads to a small but sometimes observable polarization rotation of $\mathbf{E}_\perp$, revealing contaminations from the unintended eigenmode. 
To determine $\tilde{\psi}$ for numerical eigenmodes, a calibration step is performed where a linearly polarized wave at frequency $\omega$ is launched with a constant $\mathbf{E}_\perp$, which lies along the diagonal of $y$-$z$ plane. In most cases, this wave has large overlaps with both numerical eigenmodes, so
\begin{subeqnarray}
	\label{eq:Eexpansion}
	E_y &=& E_R\sin\tilde{\psi}_R \sin(\tilde{k}_Rx +\theta_R) +E_L\sin\tilde{\psi}_L \sin(\tilde{k}_Lx +\theta_L),\\
	E_z &=& E_R\cos\tilde{\psi}_R \cos(\tilde{k}_Rx +\theta_R) +E_L\cos\tilde{\psi}_L \cos(\tilde{k}_Lx +\theta_L),
\end{subeqnarray}
have unknown amplitudes $E_R$ and $E_L$ and numerical wavevectors $\tilde{k}_R$ and $\tilde{k}_L$ are close to their analytical values $k_R$ and $k_L$. The phases $\theta_R$ and $\theta_L$ vary deterministically with $\omega t$, but the initial phases depend on the unknown expansion coefficients when spanning the wave in terms of the eigenmodes. 
Notice that each electric-field component is of the form $E=\mathcal{E}_R(\sin\tilde{k}_Rx\,\cos b_R + \cos\tilde{k}_Rx\,\sin b_R) + \mathcal{E}_L(\sin\tilde{k}_Lx\,\cos b_L + \cos\tilde{k}_Lx\,\sin b_L)$, where $b=\theta$ for $E_y$ and $b=\theta+\pi/2$ for $E_z$. By fitting a region of the simulation data where the wave envelopes are constant, as shown by the example in figure~\ref{fig:LinearEigen}, the unknown amplitudes $\mathcal{E}_R$ and $\mathcal{E}_L$, as well as the unknown phases $\theta_R$ and $\theta_L$, can be determined by linearly regressing $\mathbf{E}_\perp(x)$ against a four-dimensional basis $(\sin\tilde{k}_Rx, \cos\tilde{k}_Rx, \sin\tilde{k}_Lx, \cos\tilde{k}_Lx)$. 
In practice, the values of $\tilde{k}_R$ and $\tilde{k}_L$ are not easily known, so they are determined numerically by minimizing the residual of linear regression using $k_R$ and $k_L$ as initial guesses. Finally, using $\mathcal{E}_{y}=E\sin\tilde{\psi}$ and $\mathcal{E}_{z}=E\cos\tilde{\psi}$ from the best fit, the polarization angles $\tilde{\psi}$ of numerical eigenmode is computed. In most cases, $\tilde{\psi}$ is close to the analytical $\psi$. However, outliers can occur in limiting cases, such as when $B_0\rightarrow 0$ where the degeneracy is week, or when $\theta_B\rightarrow90^\circ$ where the ellipticity is weak. In outlier cases, it is usually sufficient to launch the desired eigenmode using $\psi$, which is calculated using the code in \citet{Shi22}.

To verify that only the intended eigenmode is launched and to measure its transmitted amplitude, a pump-only or seed-only eigen run is performed, where the laser is launched with numerically determined polarization angle $\tilde{\psi}$ and the other laser is turned off completely. Example transverse fields $E_y(x)$ and $E_z(x)$ are plotted in figures~\ref{fig:LinearEigen}.
When the wave is not an eigenmode (figure~\ref{fig:LinearEigen}$c$), the polarization trajectory precesses. On the other hand, when the wave is close to an eigenmode, its trajectory is near the stationary polarization ellipse. In figure~\ref{fig:LinearEigen}$(f)$, a slow precession can still be observed, but the contamination from the other eigenmode is small enough that it is not a major source of error. 
Using the pump-only eigen run, the transmitted pump amplitude can be measured from data shown in figures~\ref{fig:LinearEigen}$(d)$. The transmitted amplitude is normalized by analytical wave energy coefficient to compute $a_1$, which enters $\gamma_0$ via Eq.~(\ref{eq:growthrate}). Combining with the analytical $\Gamma$, the expected growth rate is computed to compare with simulations.

In the following, we will focus on eigenmodes that are right-handed with respect to the magnetic field, which usually have the largest pump-seed coupling among the four possible polarization combinations. When $\cos\theta_B>0$, meaning that $\mathbf{B}_0$ points towards $+x$, the seed laser is right-handed with respect to its wavevector. On the other hand, since the pump laser propagates in $-x$ direction, the eigenmode is left-handed with respect to its wavevector. Another way to describe the mode selection is that at a given wave frequency $\omega$, the eigenmode with smaller $k$ will be launched. This mode has an electric-field vector that co-rotates with the gyrating electrons and is therefore more strongly coupled with the electron-dominant plasma waves, such as the Langmuir-like P wave and electron-cyclotron-like F wave.

%%%%%%%%%%%%%%%%%%%%%%%%%%%%%%%%%%%%%%%%%%%
\subsection{Extracting pump and seed envelopes}
In stimulated runs where both pump and seed eigenmodes are turned on, we need to separate their contributions from the total electromagnetic fields, which are the direct observables of PIC simulations. A plausible way of separating the waves is to use Fourier-like filters. However, since the pump and seed envelopes evolve in spacetime due to three-wave interactions, their line shapes have long tails. Attempting to filter waves in the Fourier space and then taking inverse transform often introduce spurious oscillations.  
As an alternative approach, the pump and seed are directly separated in the configuration space using both electric- and magnetic-field data. Due to Faraday's law in one spatial dimension, wave fields are related in simple ways: For a right-propagating wave, $B_y=-E_z/v$ and $B_z=E_y/v$, where $v=\omega/k>0$ is the phase velocity of the wave. On the other hand, for a left-propagating wave, $B_y=E_z/v$ and $B_z=-E_y/v$ have the opposite signs.
When the signal contains a single right $(+)$ and left $(-)$ waves at constant amplitudes, the electric fields can be decomposed as $E_y =  E^+_{y} +E^-_{y}$ and $E_z = E^+_{z} + E^-_{z}$, and the magnetic fields are $B_y = -E^+_{z}/v_+  + E^-_{z}/v_-$ and $B_z = E^+_{y}/v_+ - E^-_{y}/v_-$. Notice that the right and left waves likely have different phase velocities $v_+$ and $v_-$. 
After removing static components of $\mathbf{B}_0$, the wave fields satisfy
\begin{subeqnarray}
	\label{eq:ERL}
	E^+_{y} &=& \frac{E_y+v_- B_z}{1+v_-/v_+},\quad E^+_{z} = \frac{E_z-v_- B_y}{1+v_-/v_+}, \\
	E^-_{y} &=& \frac{E_y-v_+ B_z}{1+v_+/v_-},\quad E^-_{z} = \frac{E_z+v_+ B_y}{1+v_+/v_-}.
\end{subeqnarray}
Reconstructions using the above Faraday filter is exact for constant-amplitude waves if $v_+$ and $v_-$ are known. However, due to numerical dispersion and PIC noise, $v_\pm$ differ slightly from analytical values, causing leakage through the Faraday filter.

\begin{figure}
	\centerline{\includegraphics[width=0.85\textwidth]{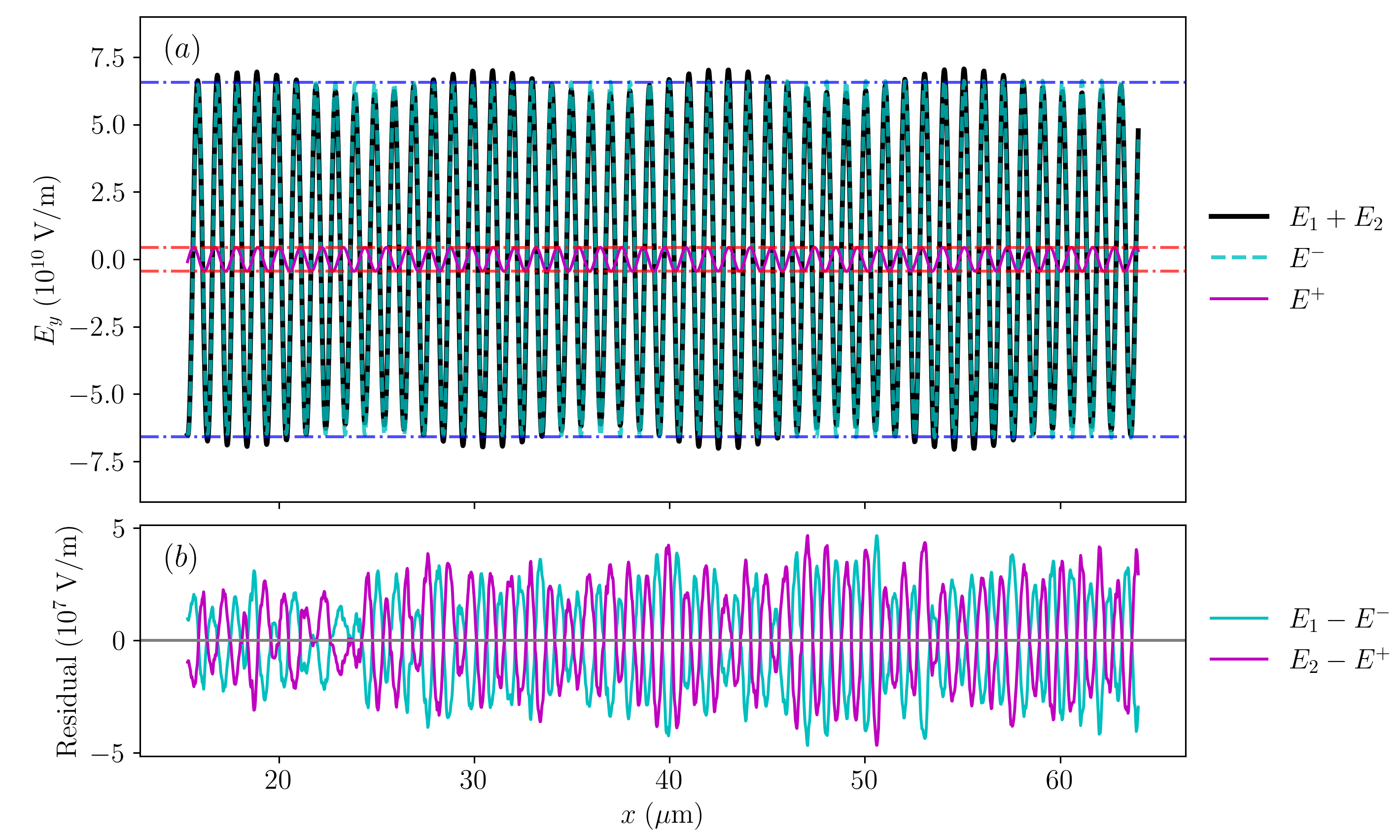}}% Images in 100% size
	\caption{$(a)$ The optimal phase velocities for separating the right ($+$) and left ($-$) propagating waves are fitted using calibration data (black), which is the sum of a pump-only ($E_1$) and a seed-only ($E_2$) runs before the laser reaches the opposite plasma boundary. The amplitudes of the pump (dashed blue) and seed (dashed red) are known, and the fitting minimizes the total difference between known and extracted pump (cyan) and seed (magenta) amplitudes. 
	$(b)$ The best-fit residuals are orders of magnitude smaller than the signals. However, the residuals have coherent leakages that are a few times larger than the PIC noise level. Fitting results for other field components are similar. The intensities are \mbox{$I_1=10^{15}\;\mathrm{W/cm}^2$} and \mbox{$I_2=5\times 10^{12}\;\mathrm{W/cm}^2$}.
	}
	\label{fig:RightLeft}
\end{figure}

To reduce the leakage, a calibration step is performed to determine the numerical phase velocities. Summing data of a pump-only run with data of a seed-only run before the wave fronts reach the opposite plasma boundary, $\tilde{v}_+$ and $\tilde{v}_-$ are fitted using Eq.~(\ref{eq:ERL}). The data is selected within the plasma region using similar criteria as in figure~\ref{fig:LinearEigen}. 
The best-fit residuals are typically orders of magnitude smaller than the signals but a few times above the PIC noise level, as shown by the example in figure~\ref{fig:RightLeft}. The coherent errors are likely due to how electric- and magnetic-field data is combined. Notice that when solving Maxwell's equations using Yee-like mesh, transverse electric fields are discretized on cell boundaries whereas transverse magnetic fields are discretized on cell centers. Therefore, interpolations are needed before $B_{i\pm1/2}$ can be summed with $E_i$. Here, a simple linear interpolation is used to estimate $B_{i}\approx (B_{i-1/2} + B_{i+1/2})/2$, which is likely the leading cause of coherent errors. Using interpolation schemes that better match the PIC algorithm may further reduce the coherent error.

However, other leakage sources are more important when the calibrated Faraday filter is applied to data from stimulated runs. Notice that the simple relation between $E$ and $B$ assumes plane waves with constant amplitudes. When the amplitude varies, additional terms arise, which depends on derivatives of the amplitude. In regimes where the amplitude varies slowly, these additional terms are small compared to the leading term but are large compared to the error from interpolation. 
Moreover, due to reflection and spontaneous scattering in PIC simulations, waves other than the pump and seed are present in the system, leading to even larger errors. For example, the pump laser reflects from the left plasma-vacuum boundary and propagates towards right together with the seed laser. 
Typically, the reflected pump amplitude is $\sim0.5\%$ of the incident amplitude, which is $\sim7\%$ of the seed amplitude when $I_1/I_2=200$. The Faraday filter is far from ideal for separating co-propagating waves, especially when they have comparable phase velocities. The leakage leads to spurious oscillations of the seed envelope, which are nevertheless usually less severe than caused by Fourier filters.

\begin{figure}
	\centerline{\includegraphics[width=0.9\textwidth]{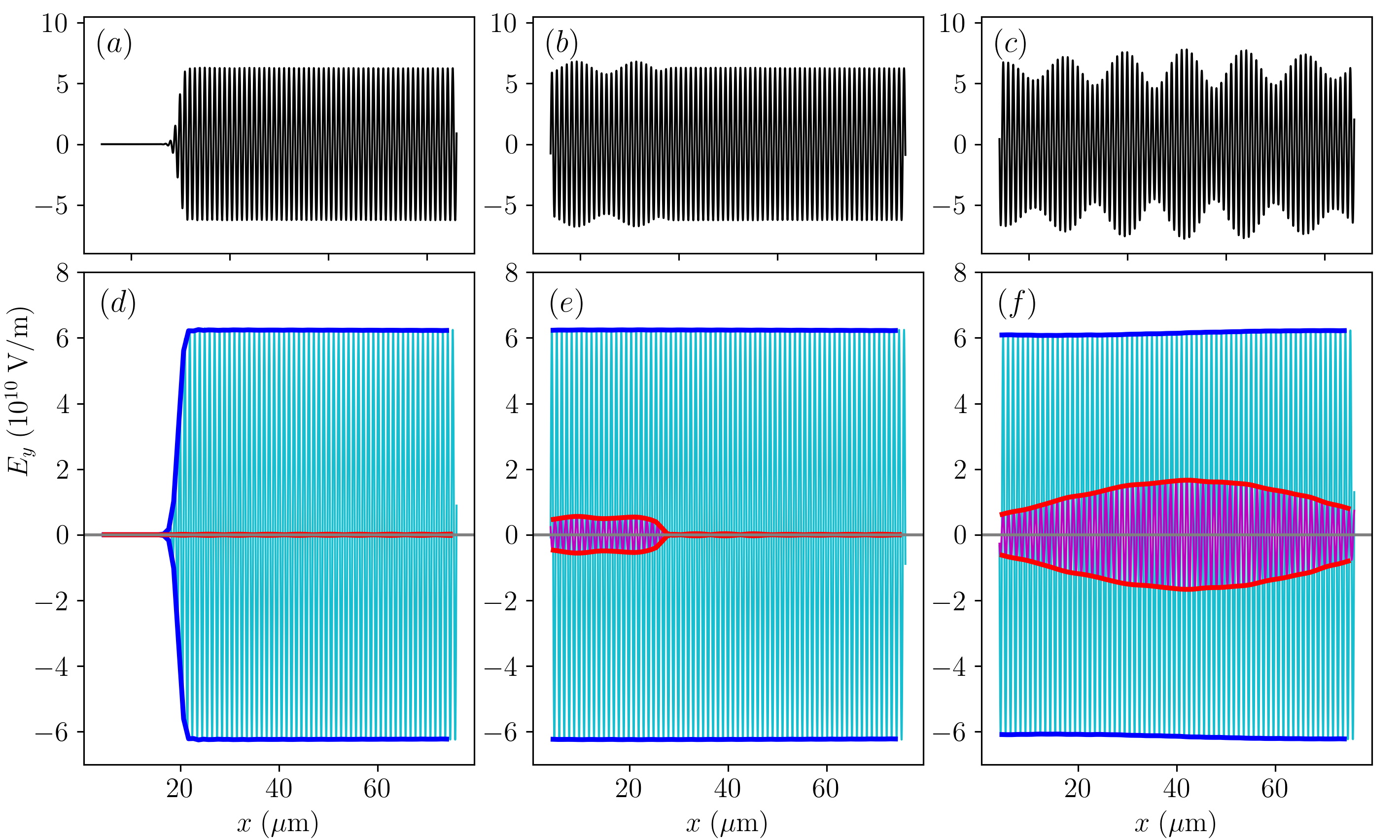}}% Images in 100% size
	\caption{Raw data $(a)$-$(c)$ is separated using a Faraday filter into pump (cyan) and seed (magenta) contributions $(d)$-$(f)$, and envelopes of pump (blue) and seed (red) are extracted using a lock-in scheme. At \mbox{$t\approx0.25$ ps} $(a, d)$, the pump almost reaches the left plasma boundary and the seed has not entered. At later time \mbox{$t\approx0.48$ ps} $(b, e)$, the pump fills the plasma and the seed enters from the left. At final time \mbox{$t\approx0.67$ ps} $(c, f)$, the seed exits the right boundary and experiences noticeable growth, and the pump is slightly depleted.
		In this example, \mbox{$T_0=10$ eV}, \mbox{$B_0=3$ kT}, $\theta_B=30^\circ$, \mbox{$I_1=10^{15}\;\mathrm{W/cm}^2$}, and \mbox{$I_2=5\times 10^{12}\;\mathrm{W/cm}^2$}. The interaction is mediated by a P wave and the resonant seed wavelength in vacuum is \mbox{$\lambda_2\approx1.09\;\mu$m}.
	}
	\label{fig:Envelope}
\end{figure}

Having separated out pump and seed electric fields, the next step is to extract the wave envelopes that enter the three-wave equations. For a wave with slowly-varying envelope, its electric field is $E(x)=\mathcal{E}(x)\sin(kx+b)+\epsilon(x)$, where $\epsilon$ is an error field. The envelope $\mathcal{E}(x)$ can be estimated by mixing $E(x)$ with a sinusoidal reference, which gives $E(x) \sin(kx+b') = \frac{1}{2}\mathcal{E}(x)[\cos(b-b')-\cos(2kx+b+b')]+\epsilon(x)\sin(kx+b')$. Averaging the mixed signal over $\lambda=2\pi/k$ ideally eliminates the last two terms to give
\begin{equation}
	\label{eq:Lock-in}
	\big\langle E(x) \sin(kx+b')\big\rangle_\lambda=\frac{1}{2}\mathcal{E}(x)\cos(b-b').
\end{equation}
The reference phase $b'$ is scanned to maximize the norm of the average, which is attained when the phases are matched $b'=b$ up to integer multiples of $\pi$. Compared to a simple moving average, the above lock-in scheme reduces sensitivities to PIC noise and coherent errors.
The numerical wave vector $\tilde{k}$, which is determined when fitting Eq.~(\ref{eq:Eexpansion}), is used to generate the sinusoidal reference. Moreover, to better remove the unwanted terms using the $\lambda$ average, a matching grid $\tilde{x}_i$ with spacing $\lambda/n$ is used, where the integer $n=\lceil\lambda/\Delta x\rceil$ and $\Delta x$ is the original spacing of the PIC grid $x_i$. The reference is generated on grid $\tilde{x}_i$ and mixed with $E(\tilde{x}_i)$, which is estimated from $E(x_i)$ using linearly interpolation. 
Averaging over $\lambda$ down samples the data, so the envelope $\mathcal{E}(x)$ lives on a grid that is more sparse than the $\tilde{x}_i$ grid by a factor of $n$. Notice that due to their wavelengths difference, the pump and seed envelopes live on two separate grids. %Nevertheless, all components of the same wave live on the same grid. 
In addition to the transverse components, the $\mathcal{E}_x$ component is estimated using the lock-in scheme alone, without attempting to separate right and left contributions. The longitudinal component is small but nonzero when lasers propagate obliquely with $\mathbf{B}_0$. Combining all components, the total scalar wave envelope $\mathcal{E}=|\mathbfcal{E}|$ is determined. 
Typical results of envelope extraction are shown in figure~\ref{fig:Envelope} for the $E_y$ component. Results are similar for the $E_z$ component, but are more noisy for the much smaller $E_x$ component. The procedure successfully separates out the pump and seed lasers, and identifies their envelopes that enclose fast-oscillating fields.

%%%%%%%%%%%%%%%%%%%%%%%%%%%%%%%%%%%%%%%%%%%
\subsection{Fitting data to analytical solutions}
\begin{figure}
	\centerline{\includegraphics[width=0.9\textwidth]{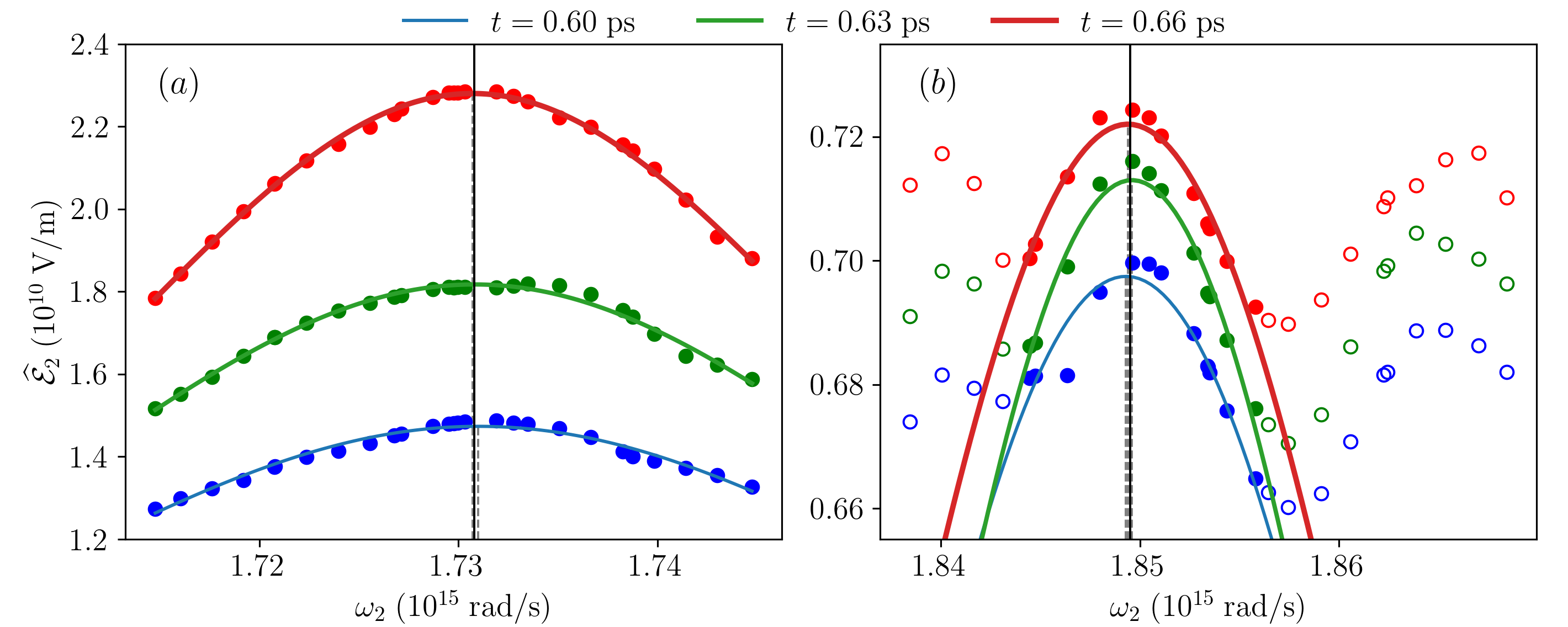}}% Images in 100% size
	\caption{The spatial maximum of seed envelope $\widehat{\mathcal{E}}_2$ is fitted to seed frequency $\omega_2$ using a Lorentzian profile to determine the resonant seed frequency. $(a)$ When \mbox{$B_0=3$ kT} and the interaction is mediated by a P wave, the data matches a simple Lorentzian, and fittings using $\widehat{\mathcal{E}}_2$ at different time slices yield consistent results (dashed vertical lines), which is averaged to give a final estimate of $\omega_2^*$ (solid vertical line). This is representative of what happens in most cases. $(b)$ When \mbox{$B_0=200$ T} and the interaction is mediated by an F wave, the data reveals multiple weak resonances. Data away from the expected warm-fluid peak are excluded from fitting (empty circles). The included data (solid circles) again yields consistent results across multiple time slices. 
		In these examples, \mbox{$T_0=10$ eV}, $\theta_B=30^\circ$, \mbox{$I_1=10^{15}\;\mathrm{W/cm}^2$}, and \mbox{$I_2=5\times 10^{12}\;\mathrm{W/cm}^2$}. 
	}
	\label{fig:FitResonance}
\end{figure}

Before fitting extracted seed envelopes to analytical solutions, we need to ensure that the seed frequency is on resonance. Notice that Eq.~(\ref{eq:three-wave}) is for resonant interactions that satisfy energy-momentum conservation $k^\mu_1=k^\mu_2+k^\mu_3$ where $k^\mu$ is the wave 4-momentum. When phase matching conditions are not satisfied, the waves can still interact, but a detuning term $\exp(ix^\mu\delta k_\mu)$ will appear in the three-wave equations. Analytical solutions in the detuned case are different from what is discussed in Sec.~\ref{Sec:Analytic}, and usually exhibit less seed growth as a result of phase modulations. 
To determine the resonant $\omega_2^*$, the seed frequency is scanned near analytically expected resonances to maximize the growth of seed envelope $\mathcal{E}_2$ in kinetic simulations. To reduce the requisite number of runs, only a few frequencies are simulated and $\widehat{\mathcal{E}}_2(\omega_2)\propto [(\omega_2-\omega_2^*)^2+\mu^2]^{-1}$ is fitted using a Lorentzian profile. Here, $\widehat{\mathcal{E}}_2$ is the spatial maximum of $\mathcal{E}_2$ at a given time slice and $\mu$ is a phenomenological linewidth.
In most cases, the fitting behaves as expected, as shown in figure~\ref{fig:FitResonance}$(a)$. Using $\widehat{\mathcal{E}}_2$ at different time slices yields consistent estimates of $\omega_2^*$ (dashed vertical lines), as long as $\mathcal{E}_2$ has grown substantially above the noise level. In most cases, only nine $\omega_2$ values are scanned within a $3\times10^{13}$-rad/s window near the expected resonance. The fitting uses last four simulation outputs, when the seed has propagated across more than half the plasma length. The four estimates of $\omega_2^*$ are averaged using equal weight to give a final estimate (solid vertical line). 
However, in some cases, more complicated spectra emerge during the scan, as shown in figure~\ref{fig:FitResonance}$(b)$. In this example, \mbox{$B_0=200$ T} and the electron gyro frequency is a few times smaller than the plasma frequency. In this regime, the kinetic wave dispersion relation involves multiple Bernstein-like waves near the expected warm-fluid resonance. In cases like this, additional $\omega_2$ values are scanned to resolve individual peaks, and only data near the expected resonance (solid circles) is used to fit $\omega_2^*$. Using data at multiple time slices again yields consistent estimates. 
As an intriguing observation, spacings between the side peaks do not seem to have a simple dependence on $B_0$ and $\theta_B$, and often appears to change with time. These features remain to be investigated in the future.

\begin{figure}
	\centerline{\includegraphics[width=0.72\textwidth]{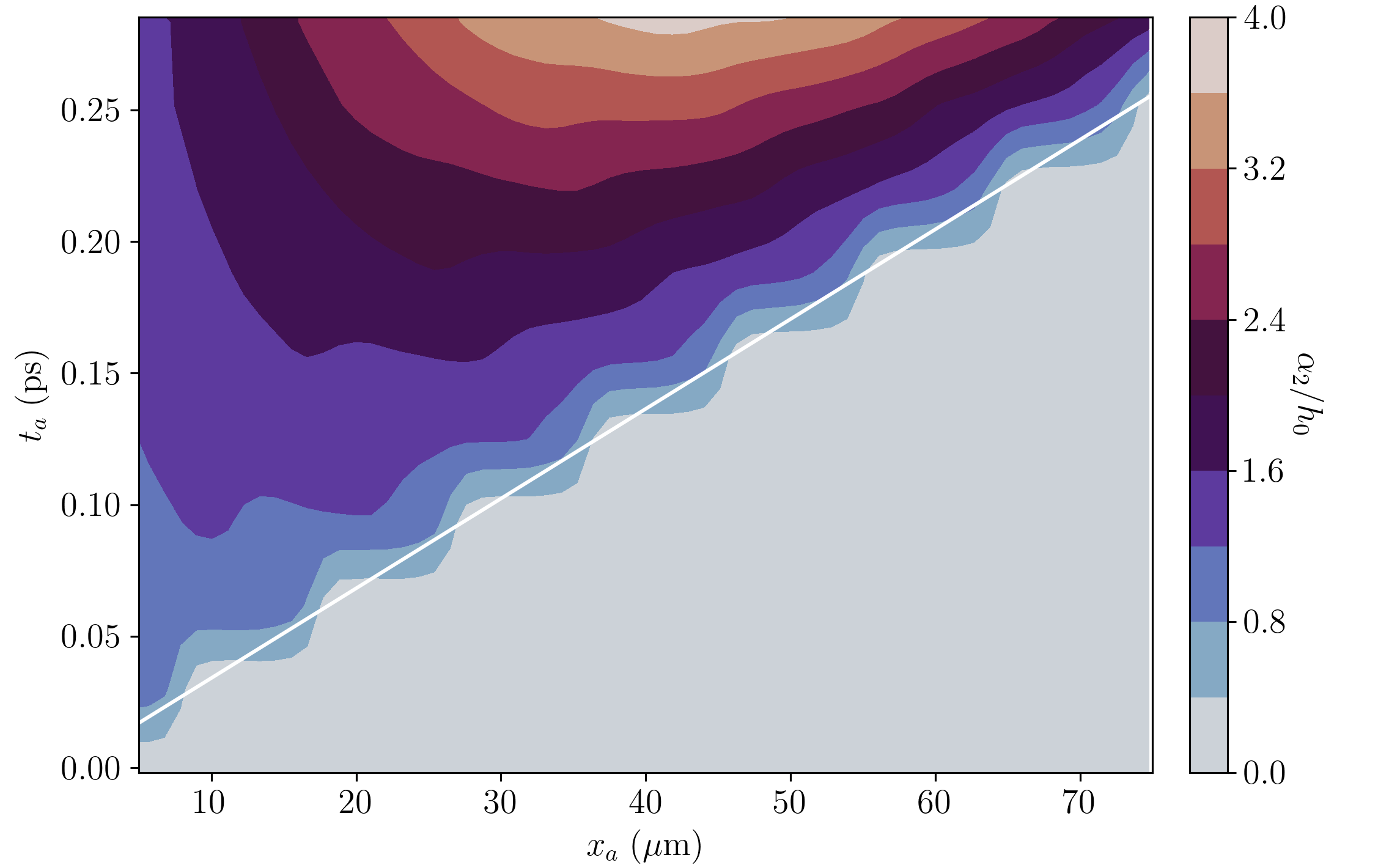}}% Images in 100% size
	\caption{Normalized seed amplitude $\alpha_2/h_0$ in the analytical frame $(x_a, t_a)$. The step-like feature near the wave front is a visualization artifact when only nine time slices are plotted. The wave front $x_a=\tilde{v}_2 t_a$ (white line) is fitted to determine the numerical group velocity $\tilde{v}_2$. This example is the same run as in figure~\ref{fig:Envelope}.
	}
	\label{fig:WaveFront}
\end{figure}

Using resonant seed to setup stimulated runs, the extracted seed envelopes are placed into analytic frame to prepare for fitting, as shown in figure~\ref{fig:WaveFront}. 
Notice that in the step-function problem, the plasma-vacuum boundary is sharply at $x=0$ and a constant-amplitude seed arrives at the boundary when $t=0$. These idealized setups are necessarily softened in simulations, where the plasma expands into the vacuum forming a sheath region, and the seed is ramped up using a $\tanh$ profile whose temporal width equals to the seed period to smooth out numerical artifacts. 
For the spatial axis, to avoid boundary effects, data within $4\lambda_1\gg\lambda_D$ from the initial plasma-vacuum boundaries are excluded, where $\lambda_D$ is the Debye length. Hence, the data frame $x_d$ is shifted from the analytic frame $x_a=x_d+x_0$ by an offset $x_0$.  
For the time axis, since the seed is launched after the pump has propagated across the plasma, the data frame $t_d$ is delayed from the analytic frame $t_a=t_d-t_0$ for some offset $t_0$. 
To place data into the analytic frame, the seed wave front $x_a=\tilde{v}_2t_a$ is fitted to determine both offsets and the numerical group velocity $\tilde{v}_2$, which is close to but not equals to the analytical group velocity $v_2=\partial\omega/\partial k_x$ due to numerical dispersions. 
The seed wave front is identified as the location where the seed envelope $\mathcal{E}_2$ drops below half its boundary value, and the thickness of the wave front is identified as the spatial separation where $\mathcal{E}_2$ attains $10\%$ and $90\%$ of its boundary value. Only data behind the wave front by twice its thickness are used for fitting. 
Finally, we need to normalize the electric field envelope $\mathcal{E}_2$. Notice that analytical solution of the step-function problem can be written in terms of the ratio $\alpha_2/h_0$, where $h_0$ is the boundary value. In simulations, the value at the domain boundary is an input, but the value at the plasma-vacuum boundary is not controlled, because only a fraction of the incident seed laser is transmitted into the plasma. 
The boundary value inside the plasma is determined using the seed-only eigen run. Using calibration data similar to what is shown in figure~\ref{fig:LinearEigen}$(d)$, the transmitted seed amplitude is measured and used to normalize envelopes in the stimulated run. Notice that the calibration needs to be performed each time plasma conditions are changed or the seed laser is varied.

\begin{figure}
	\centerline{\includegraphics[width=0.7\textwidth]{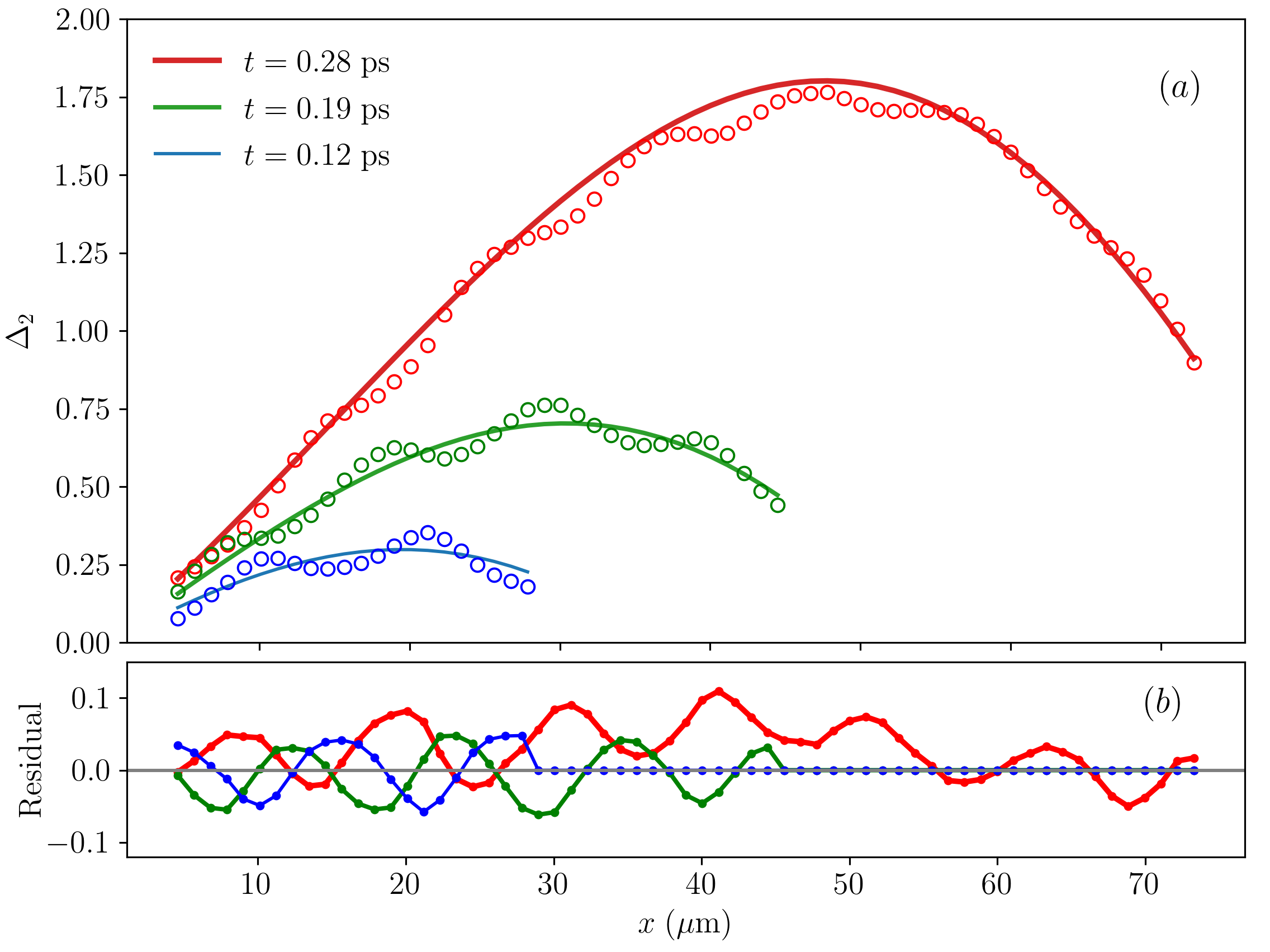}}% Images in 100% size
	\caption{$(a)$ The growth function extracted from simulation data (circles) is well matched by the analytical solutions (lines) using best-fit parameters. Results are shown for three representative time slices as the seed propagates and grows. The fitting quality at other time slices are similar. 
		$(b)$ Residuals of best fit are an order of magnitude smaller than the signal, and are dominated by leakage during pump-seed separation. The reflected pump amplitude is $\sim7\%$ of the seed.
		In this example, \mbox{$T_0=100$ eV}, \mbox{$B_0=3$ kT}, $\theta_B=30^\circ$, \mbox{$I_1=10^{15}\;\mathrm{W/cm}^2$}, \mbox{$I_2=5\times 10^{12}\;\mathrm{W/cm}^2$}, and the interaction is mediated by a P wave. The best-fit parameters are \mbox{$\tilde{v}_2\approx2.97\times10^8$ m/s}, \mbox{$\tilde{\mu}_3\approx5.17\times10^{12}$ rad/s}, and \mbox{$\tilde{\gamma}_0\approx9.25\times10^{12}$ rad/s}.
	}
	\label{fig:FitEquation}
\end{figure}

Having extracted $\alpha_2/h_0$ from the stimulated run, the data in analytic frame is fitted to Eq.~(\ref{eq:DeltaStep}). 
Here, the simplification $v_3=0$ is made because $|v_3|\ll v_2$. From figure~\ref{fig:Delta_spatial}, we see that $v_3=0$ is already a good approximation even when $v_3=-v_2/10$. For a warm-fluid plasma with moderate temperature \mbox{$T_0\sim10$ eV}, the ratio $v_3/v_2\sim O(10^{-2})$ is much smaller, so the limiting-form growth function provides an even better approximation. Evaluating the growth function requires a single numerical integral, which is much cheaper than computing the double integral in Eq.~(\ref{eq:alphaStep}) that gives the exact solution at finite $v_3<0$. 
As another simplification, the physical collision module is turned off in PIC simulations, so lasers are undamped beyond spurious numerical collisions. Setting $\mu_2=0$ gives $\Delta_2=\alpha_2/h_0-1$. The right-hand side of this expression is determined from simulation data to give $\widetilde{\Delta}_2$, and the left-hand side is the analytical formula, which contains three parameters $v_2$, $\mu_3$, and $\gamma_0$. Since $\tilde{v}_2$ is already determined from fitting the wave front, only two parameters remain to be fitted. 
The fitting is treated as an optimization problem, where parameters are searched near their expected values to minimize the residual $||\widetilde{\Delta}_2-\Delta_2||$, where $||f|| = \frac{1}{n_xn_t} [\sum_{i=1}^{n_x}\sum_{j=1}^{n_t}|f(x_i,t_j)|^2]^{1/2}$. The discrepancy $\widetilde{\Delta}_2-\Delta_2$ is set to zero outside the region where data is valid. As discussed earlier, the data is valid when $x$ is sufficiently far away from both the plasma boundaries and the seed front, and when $t$ is after seed experiences noticeable growth but $a_1(x)$ and $f_e(v_x)$ remains largely unchanged. 
Typical best-fit results are shown in figure~\ref{fig:FitEquation}, where the simulation data (circles) is well matched by analytical solutions (lines), and the residuals are dominated by leakages during pump-seed separation.

\begin{figure}
	\centerline{\includegraphics[width=0.9\textwidth]{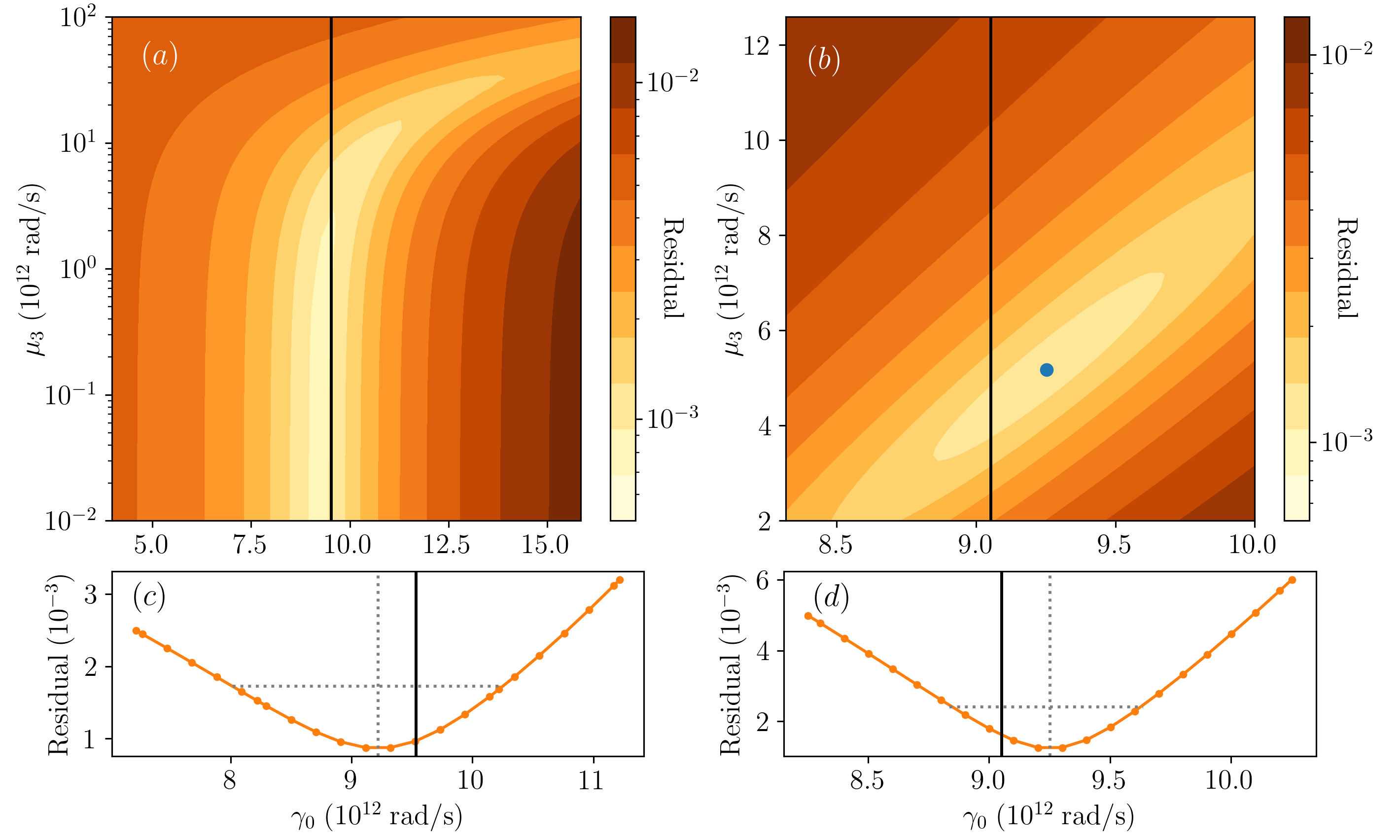}}% Images in 100% size
	\caption{Uncertainties of fitting parameters are quantified by scanning the fit residual. 
		$(a, c)$ When \mbox{$T_0=10$ eV} is cold, damping is weak and poorly constrained. The best fit is far below the vertical scale and is consistent with a negligible $\mu_3$. Nevertheless, $\gamma_0$ is well constrained and matches analytical growth rate (vertical solid line).
		$(b, d)$ When \mbox{$T_0=100$ eV} is hotter, damping becomes stronger. In this case, both $\gamma_0$ and $\mu_3$ are well constrained. The best fit (blue dot) matches the analytical growth rate within error bar, which is defined as the range where fit residual doubles its minimum. 
		$(c, d)$ Using the best-fit $\mu_3$, one dimensional scan along $\gamma_0$ gives marginal fit residuals and error bars (horizontal dashed line). 
		In these examples, \mbox{$B_0=3$ kT}, $\theta_B=30^\circ$, \mbox{$I_1=10^{15}\;\mathrm{W/cm}^2$}, \mbox{$I_2=5\times 10^{12}\;\mathrm{W/cm}^2$}, and interactions are mediated by P waves. $(b)$ and $(d)$ are the same run as in figure~\ref{fig:FitEquation}. 
	}
	\label{fig:r0-mu3_scan}
\end{figure}

To quantify uncertainty, fit residual is scanned over a range of parameter values to estimate error bars. As parameters move away from their best-fit values, the fit residual increases. Error bars are defined as the parameter range beyond which the residual doubles its minimum. Using this definition, in the limit where data exactly matches analytical solutions, the residual is zero and the error bar is also zero. In the opposite limit where simulation and theory poorly match, the residual is large and the error bar is wide. 
Two examples are shown in figure~\ref{fig:r0-mu3_scan}, where the resonant interactions are mediate by Langmuir-like P waves. %For both examples, \mbox{$B_0=3$ kT}, $\theta_B=30^\circ$, \mbox{$I_1=10^{15}\;\mathrm{W/cm}^2$}, and \mbox{$I_2=5\times 10^{12}\;\mathrm{W/cm}^2$}. 
When the temperature \mbox{$T_0=10$ eV} is cold (figure~\ref{fig:r0-mu3_scan}$a$), collisionless damping of the plasma wave is miniscule. In this case, the residual remains small for a wide range of $\mu_3$ values, and the best fit is \mbox{$\tilde{\mu}_3\approx1.36\times10^{-2}$ rad/s}, which is consistent with zero damping. On the other hand, the residual increases rapidly when $\gamma_0$ deviates from its best-fit value \mbox{$\tilde{\gamma}_0\approx9.22\times10^{12}$ rad/s}, which is close to the analytical value \mbox{$\gamma_0^a\approx9.53\times10^{12}$ rad/s} (solid vertical line). The marginal error bar is shown in figure~\ref{fig:r0-mu3_scan}$(c)$, where $\mu_3$ is fixed at its best-fit value and $\gamma_0$ is scanned. The residual is convex near $\tilde{\gamma}_0$ (dashed vertical line), and $\gamma_0^a$ is within the error bar (dashed horizontal line). 
When \mbox{$T_0=100$ eV} is hotter (figure~\ref{fig:r0-mu3_scan}$b$), collisionless damping is stronger, so both $\gamma_0$ and $\mu_3$ are constrained. As discussed earlier, magnetized collisionless damping is not easy to compute, so the procedure here provides a unique method for measuring the damping rate, which in this case is best fitted by \mbox{$\tilde{\mu}_3\approx5.17\times10^{12}$ rad/s}.
The best-fit \mbox{$\tilde{\gamma}_0\approx9.25\times10^{12}$ rad/s} matches the analytical value \mbox{$\gamma_0^a\approx9.05\times10^{12}$ rad/s} within the marginal error bar shown in figure~\ref{fig:r0-mu3_scan}$(d)$. 
The marginal residual is always convex and usually skewed towards smaller $\gamma_0$. 
The two-dimensional landscape of the residual is flat along a valley in the $\gamma_0$-$\mu_3$ space. This is intuitive because a larger damping cancels the effect of a larger growth. In the asymptotic limit $t\rightarrow+\infty$, the exponential solution $e^{\kappa x}$ only depends on the ratio $\kappa=\gamma_0^2/\mu_3v_2$, so $\gamma_0$ and $\mu_3$ are not independently constrained. In other words, the separate constrains come from solutions at earlier time. As discussed after Eq.~(\ref{eq:DeltaStep}), the growth function $\Delta_2$ is symmetric about $x=\frac{1}{2}v_2t$ in the absence of damping. As $\mu_3$ increases, the growth behind $x=\frac{1}{2}v_2t$ is reduced more, leading to an asymmetric profile of $\Delta_2$. The asymmetry in the transient-time profile is what allows the effects of $\mu_3$ and $\gamma_0$ to be distinguished.

%%%%%%%%%%%%%%%%%%%%%%%%%%%%%%%%%%%%%%%%%%%
\subsection{Scanning physical parameters\label{Sec:Results}}
\begin{figure}
	\centerline{\includegraphics[width=0.93\textwidth]{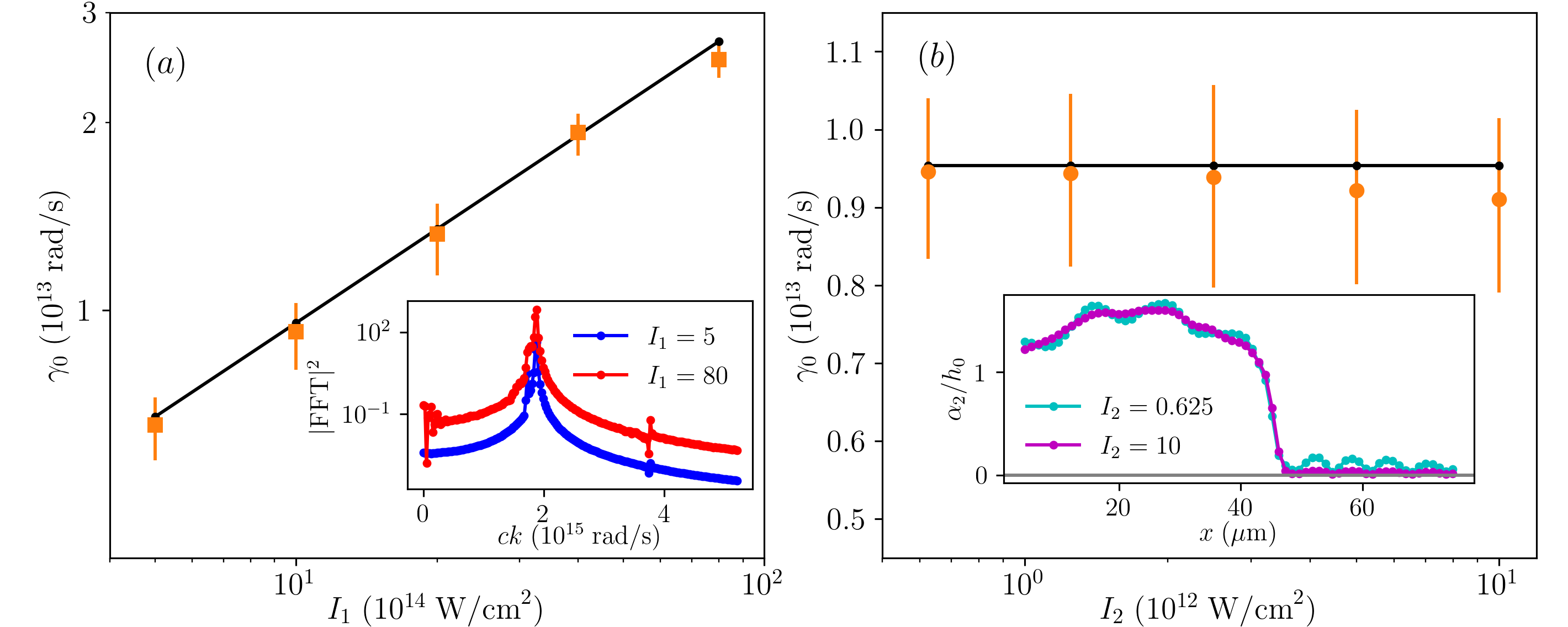}}% Images in 100% size
	\caption{The bare growth rate $\gamma_0\propto I_1^{1/2}$ scales with pump intensity $(a)$ and is independent of the seed intensity $(b)$ within the range where simulations are described by the linearized three-wave equations. Theory predictions (black lines) match simulation results (orange symbols) within error bars.
	At large pump intensity $I_1$, spontaneous scattering depletes the pump and excites additional waves, as shown in the inset of $(a)$ where the Fourier power spectra are plotted for transverse electric fields within the plasma region.
	At larger seed intensity $I_2$, pump depletion occurs earlier, which reduces the amount of data that can be used for fitting analytical solutions. On the other hand, at smaller $I_2$, the leakage during pump-seed separation is more detrimental, as shown in the inset of $(b)$ where the normalized envelopes $\alpha_2/h_0$ suffer from spurious oscillations. 
	In this set of scans, \mbox{$T_0=10$ eV}, \mbox{$B_0=3$ kT}, $\theta_B=30^\circ$, the vacuum seed wavelength is \mbox{$\lambda_2\approx1.088\;\mu$m}, and the interaction is mediated by a P wave. 
	}
	\label{fig:Iscan}
\end{figure}

Having described in detail how simulations are set up and how data is processed, the protocol is applied to scan physical parameters and compare simulations with theory. First, it is useful to verify that the pump and seed intensities are selected properly. As discussed at the beginning of this section, the pump and seed intensities are constrained by \mbox{$10^{10}\;\mathrm{W/cm}^2$} $\ll I_2\sim10^2 I_1\ll$ \mbox{$10^{14}\;\mathrm{W/cm}^2$}. Within this range, the bare growth rate $\gamma_0\propto a_1\propto I_1^{1/2}$ scales with the pump intensity $I_1$ but is independent of $I_2$. This theory prediction is confirmed by results shown in figure~\ref{fig:Iscan}, where black lines are theory and orange symbols with error bars are simulations. 
In figure~\ref{fig:Iscan}$(a)$, the seed intensity is fixed at \mbox{$I_2=5\times10^{12}\;\mathrm{W/cm}^2$} and the agreement is excellent within the range $I_1$ is scanned. Notice that the black line is not a fit of data.
A hint of deviation can be seen from the last data point at \mbox{$I_1=8\times10^{15}\;\mathrm{W/cm}^2$}, where data starts to drop below the theory line. This behavior is expected because pump that is too intense causes large spontaneous scattering that depletes the pump. 
Signatures of spontaneous scattering are clearly visible in the inset, which shows the Fourier power spectrum of transverse electric fields within the plasma region. At \mbox{$I_1=5\times10^{14}\;\mathrm{W/cm}^2$}(blue), the spectrum is clean and dominated by the pump and seed near \mbox{$ck=2\times10^{15}$ rad/s} and the plasma wave near \mbox{$ck=4\times10^{15}$ rad/s}. In comparison, at \mbox{$I_1=8\times10^{15}\;\mathrm{W/cm}^2$}(red), extra waves are excited due to spontaneous scattering. These extra waves remove energy that would otherwise be used to grow the seed, and they also constitute a messy background that jeopardizes data analysis. 
In figure ~\ref{fig:Iscan}$(b)$, the agreement remains excellent within the range $I_2$ is scanned, where the pump intensity is fixed at \mbox{$I_1=10^{15}\;\mathrm{W/cm}^2$}. The error bars remain largely constant due to two competing effects: At stronger seed intensities, pump depletion occurs earlier so more data is excluded from fitting, resulting in less constraints and larger error bars. On the other hand, at weaker seed intensities, the leakage during pump-seed separation more severely distorts the seed profile, causing poorer fits and larger error bars. The two competing effects cause error bars to remain roughly constant within this $I_2$ range. Effects of leakage can be clearly seen in the inset. The normalized envelope $\alpha_2/h_0$ is relatively smooth when \mbox{$I_2=10^{13}\;\mathrm{W/cm}^2$} (magenta), but the spurious oscillations become more severe when \mbox{$I_2=6.25\times10^{11}\;\mathrm{W/cm}^2$} (cyan). The artifacts are primarily due to reflected pump that co-propagates with the seed, which cannot be separated using Eq.~(\ref{eq:ERL}) and Eq.~(\ref{eq:Lock-in}), causing spurious oscillations. 
Limited by these effects, remaining simulations use \mbox{$I_1=10^{15}\;\mathrm{W/cm}^2$} and \mbox{$I_2=5\times10^{12}\;\mathrm{W/cm}^2$}.

\begin{figure}
	\centerline{\includegraphics[width=0.7\textwidth]{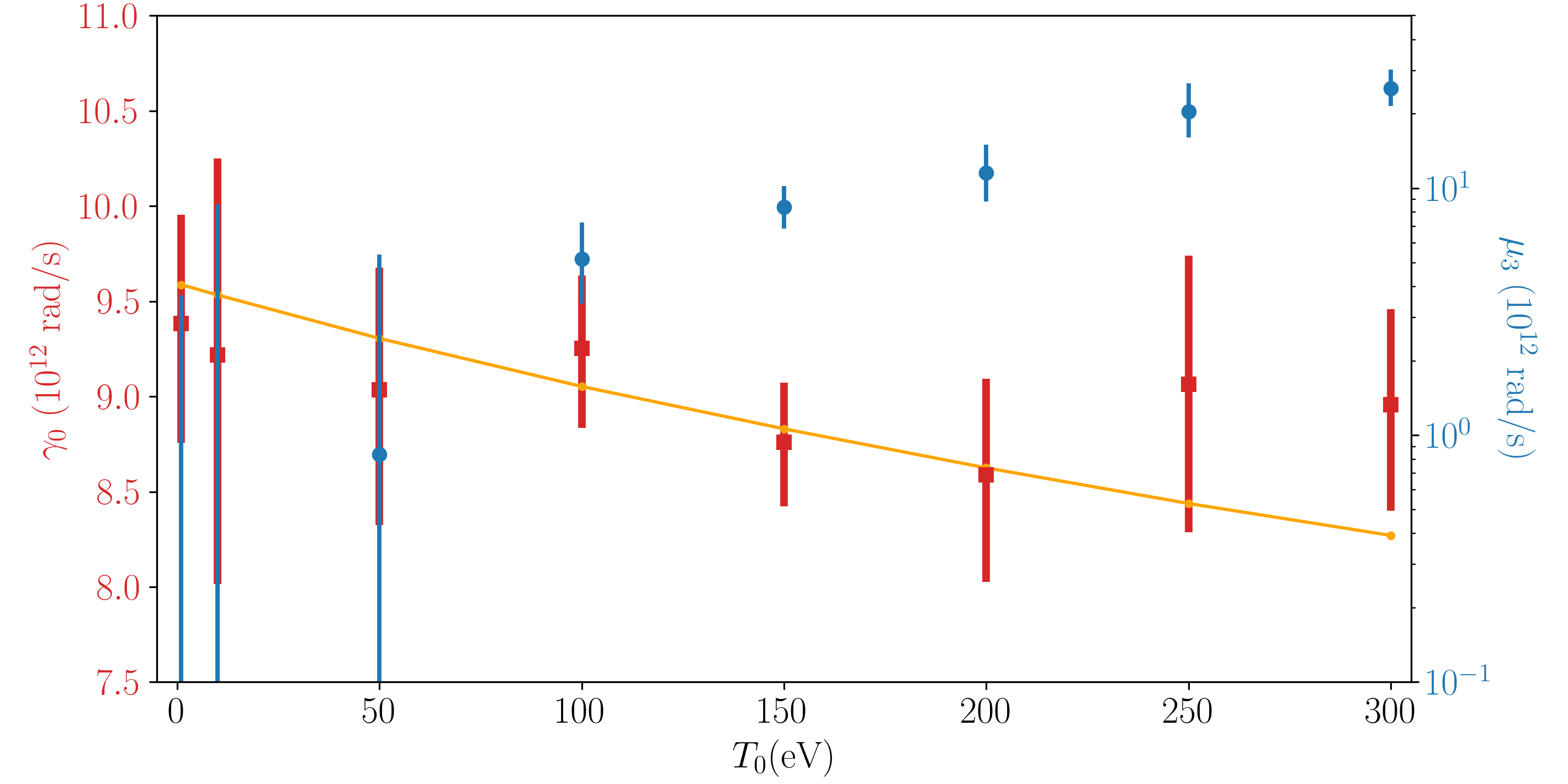}}% Images in 100% size
	\caption{The growth rate $\gamma_0$ (left axis) and damping rate $\mu_3$ (right axis) as functions of the the plasma temperature $T_0$. Colored symbols with error bars are simulation results, and the orange line is the expected growth rate in warm-fluid theory. 
		When $T_0$ increases by tenfold, $\mu_3$ increases by orders of magnitude. On the other hand, $\gamma_0$ decreases only slightly, and data is well matched by theory until the last two points.
		In this set of scans, \mbox{$B_0=3$ kT}, $\theta_B=30^\circ$, \mbox{$I_1=10^{15}\;\mathrm{W/cm}^2$}, \mbox{$I_2=5\times10^{12}\;\mathrm{W/cm}^2$}, and the interactions are mediated by P waves. 
	}
	\label{fig:Tscan}
\end{figure}

Second, the protocol is used to scan plasma temperature $T_0$ to determine the range of validity of warm-fluid theory (figure~\ref{fig:Tscan}).
At low temperature, collisionless damping of plasma waves are negligible and the $\mu_3$ error bars are large. 
On the other hand, when \mbox{$T_0>10^2$ eV}, damping starts to dominate growth and data start to deviate from warm-fluid theory. At \mbox{$T_0\sim10^3$ eV}, the seed barely grows unless the pump is stronger. However, a stronger pump suffers more from spontaneous scattering, leading to systematic errors and large uncertainties. Therefore, the applicability of the protocol does not far exceed the temperature range reported here. 
Within this range, the expected $\gamma_0$ from warm-fluid theory (line) agrees with simulation data (squares) up to \mbox{$T_0=300$ eV}, where hints of disagreements emerge. It is remarkable that even when collisionless damping has become significant, which is a purely kinetic phenomenon, the three-wave coupling remains well described by the warm-fluid theory. 
In subsequent scans, the plasma temperature is fixed at \mbox{$T_0=10$ eV} where damping is negligible.

\begin{figure}
	\centerline{\includegraphics[width=0.73\textwidth]{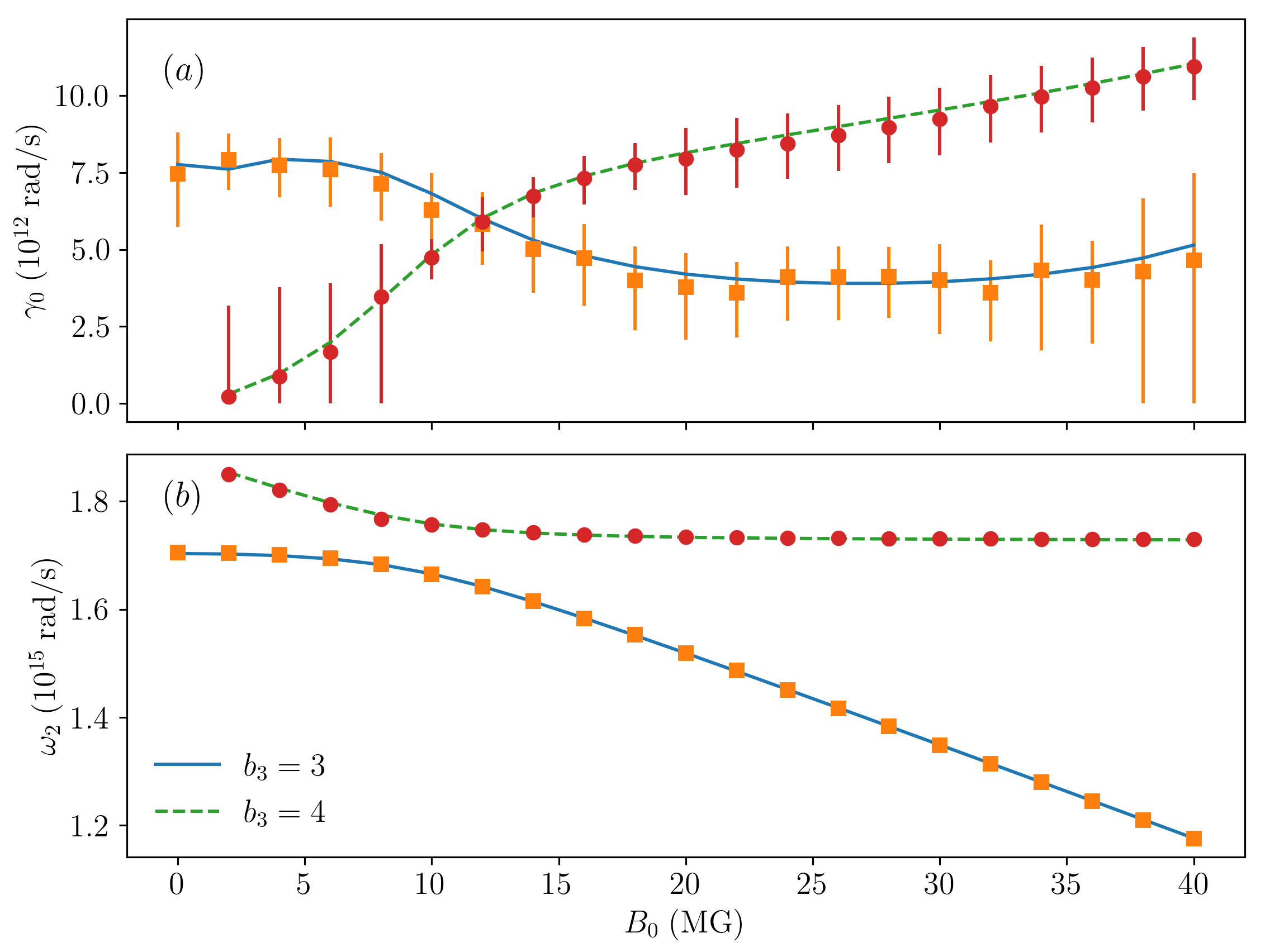}}% Images in 100% size
	\caption{$(a)$ Growth rate $\gamma_0$ and $(b)$ resonant seed frequency $\omega_2$ as functions of the background magnetic field $B_0$ when interactions are mediated by electron-dominant P and F waves. Analytical results (lines) match simulation data (symbols). Error bars are shown for $\gamma_0$ but are too small to show for $\omega_2$. The index $b_3$ labels the wave branch of $\alpha_3$. For example, $b_3=4$ means $\alpha_3$ is the 4-th highest frequency branch of the warm-fluid dispersion relation.  
		In this set of scans, \mbox{$T_0=10$ eV}, $\theta_B=30^\circ$, \mbox{$I_1=10^{15}\;\mathrm{W/cm}^2$}, and \mbox{$I_2=5\times10^{12}\;\mathrm{W/cm}^2$}. 
	}
	\label{fig:Bscan}
\end{figure}

Third, the protocol is used to scan the background magnetic field strength $B_0$ for interactions that are mediated by electron-dominant resonances (figure~\ref{fig:Bscan}).
In weak magnetic fields, the $b_3=3$ resonance, namely, the $\alpha_3$ eigenmode whose frequency is the third highest at a given wavevector, is the Langmuir-like P wave. In particular, at $B_0=0$, this resonance is precisely mediated by the Langmuir wave that gives rise to Raman backscattering in unmagnetized plasmas. When $B_0\ne0$ and $\theta_B\ne0$, this eigenmode is modified due to cyclotron motion, but its frequency only weakly dependents on $B_0$, as shown in figure~\ref{fig:Bscan}$(b)$.
Beyond \mbox{$B_0\approx10$ MG}, the eigenmodes cross over, and the $b_3=3$ resonance becomes the electron-cyclotron-like F wave, whose frequency $\omega_3$ increases almost linearly with $B_0$. 
Complementarily, the $b_3=4$ resonance is the F wave in weak $B_0$ and crosses over to become the P wave in strong magnetic fields. 
Within the $B_0$ range scanned here, the interaction mediated by the F wave provides a smaller coupling than the P wave at the same $B_0$. In particular, at $B_0=0$, the F wave vanishes and provides zero coupling. The exception is near \mbox{$B_0=10$ MG} when the two eigenmodes strongly hybridize. Although they remain two distinct eigenmodes, their couplings during frequency crossover become equal. 
Immediately after the crossover, P-wave coupling recovers its unmagnetized strength and continues to increase, while F-wave coupling drops before starting to increase again. 
In stronger fields not shown here, $\gamma_0$ continues to increase and F-wave coupling surpasses P-wave coupling \citep{shi2019amplification}. In even stronger fields, the F wave frequency approaches $\omega_1/2$ and two-magnon decays are encountered \citep{Manzo22}, for which kinetic effects are dominant. After the F wave frequency increases beyond $\omega_1/2$, it can no longer mediate resonant interactions, so only the P resonance remains. As $B_0$ further increases, mediation by P wave continues to strengthen, and a strong resonance is encountered when electron gyro frequency approaches $\omega_1$ \citep{edwards2019laser}. Beyond that point, the \mbox{1-$\mu$m} pump frequency is below the R-wave cutoff, so only L-wave pump can propagate, whose coupling is substantially weaker. 
In the range scanned in figure~\ref{fig:Bscan}, kinetic effects are not overwhelming and the agreement is excellent. Notice that the lines are not fits of data. Although error bars for $\gamma_0$ can perhaps be reduced, growth rates extracted from simulations already track detailed features of analytical predictions.

\begin{figure}
	\centerline{\includegraphics[width=0.73\textwidth]{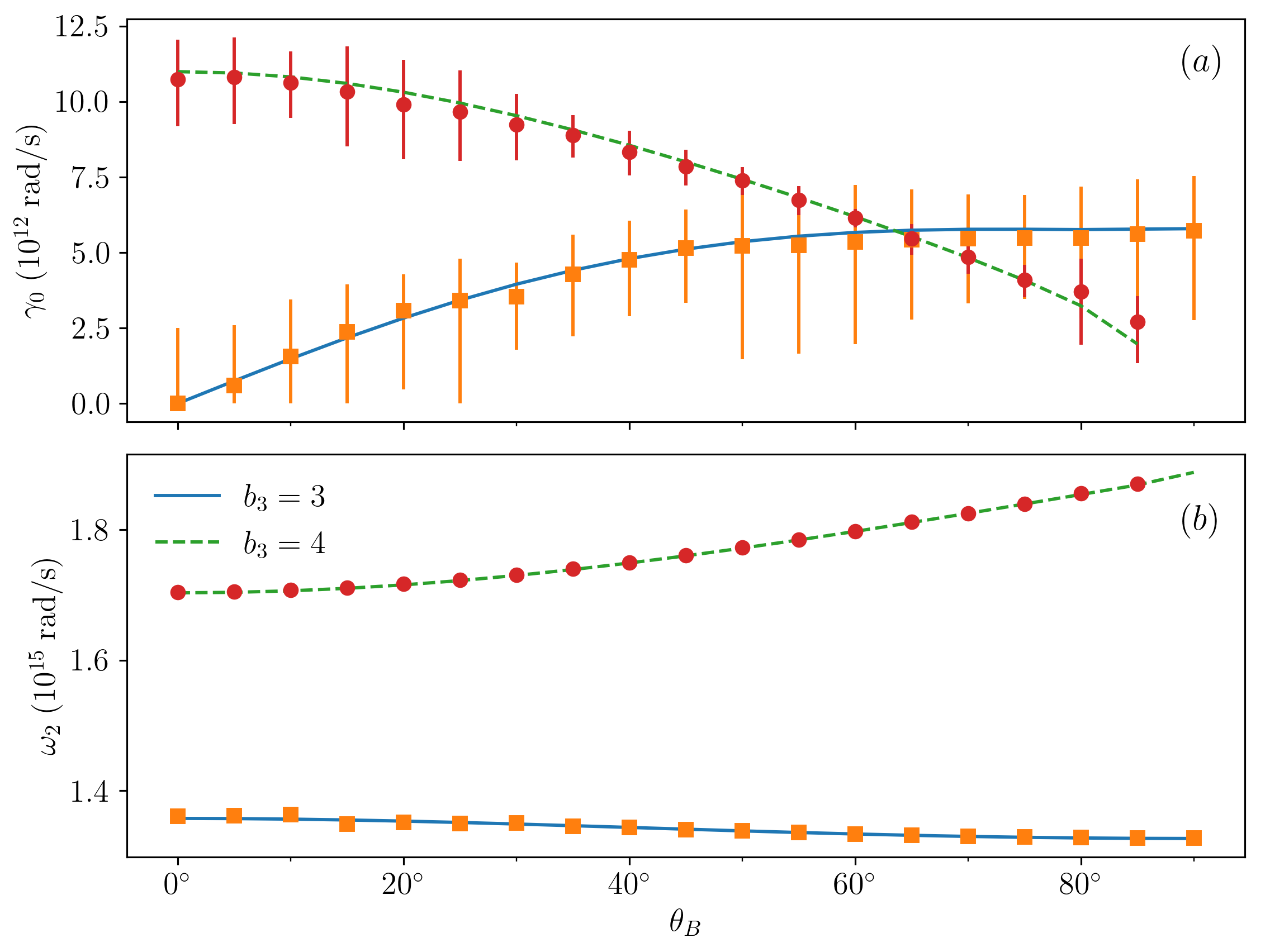}}% Images in 100% size
	\caption{Same as figure~\ref{fig:Bscan}, but for $\theta_B=\langle\mathbf{k},\mathbf{B}_0\rangle$ scan at fixed \mbox{$B_0=3$ kT}. 
	}
	\label{fig:thetaScan}
\end{figure}

Finally, the protocol is used to scan $\theta_B$, the angle between the magnetic field and the wavevectors, when interactions are mediated by the two highest-frequency resonances (figure~\ref{fig:thetaScan}). 
The magnetic field is fixed at \mbox{$B_0=3$ kT}, where the electron cyclotron frequency is larger than the plasma frequency. In this case, the $b_3=3$ resonance is the electron-cyclotron wave when $\theta_B=0$ and becomes the UH wave when $\theta_B=90^\circ$. The other resonance with branch index $b_3=4$ is the Langmuir wave when $\theta_B=0$ and becomes the LH wave when $\theta_B=90^\circ$. 
The coupling provided by the electron-cyclotron wave is zero, because the interaction is both polarization forbidden and energy forbidden \cite{shi2018laser}. The coupling provided by the LH wave is finite, but smaller than UH mediation by a factor that is roughly proportional to $(\omega_{\mathrm{LH}}/\omega_{\mathrm{UH}})^{1/2}\ll1$ \citep{Shi17}.
On the other hand, the coupling provided by the electron-dominant UH wave is strong, and has a comparable strength as unmagnetized Raman scattering.
As $\theta_B$ increases, the hybridization between cyclotron motion and electrostatic oscillation changes. The growth mediated by the $b_3=3$ resonance increases, while the growth mediated by the $b_3=4$ resonance decreases. The simulation data (symbols) and analytical results (lines) agree within error bars. Notice that the lines are not fits of data. The data point at $b_3=4$ and $\theta_B=90^\circ$ is missing because the LH coupling is too weak and the resonance cannot be unambiguously identified.

%%%%%%%%%%%%%%%%%%%%%%%%%%%%%%%%%%%%%%%%%%%
%%%%%%%%%%%%%%%%%%%%%%%%%%%%%%%%%%%%%%%%%%%
\section{Discussion\label{Sec:Discussion}}
While strong resonances can be fitted using the protocol described in this paper, weak resonances are challenging to extract. In warm-fluid plasmas, weak resonances include the LH wave, as well as sound-like waves, Alfv\'en-like waves, and ion-cyclotron-like waves. In these resonances, ion motion is important, so the coupling are typically smaller than electron-dominant modes by factors proportional to $1/\sqrt{M_i}\ll1$. Weak resonances are difficult to isolate because they compete with much stronger spontaneous scattering in PIC simulations. Within the limited time window before spontaneous scattering overwhelms simulation results, the stimulated scattering mediated by a weak resonance barely experiences any growth, and is therefore difficult to observe amidst leakages during pump-seed separation.

The difficulties may be circumvented in a number of ways. First, numerical noise can be reduced to suppress unphysical spontaneous scattering. This can be achieved using less noisy Vlasov simulations. However, Vlasov codes can be expensive when an oblique background magnetic field is present, because the simulations need to include at least one spatial dimension and three velocity directions. Alternatively, PIC simulations with a substantially larger number of particles $N$ can be used to reduce numerical fluctuations. However, this can also be expensive because PIC noise only decreases slowly with $\sqrt{N}$. 
Second, plasma parameters can be chosen in regimes where electron modes are strongly damped or chirped. For example, at keV temperature that is relevant for fusion, collisionless damping rates for electron modes are stronger than the growth rates. In comparison, due to heavy ion masses, ion-dominant modes are less damped. The balance between damping and growth can make ion-dominant modes competitive with electron-dominant modes, and therefore survive numerical artifacts. It remains to be seen whether couplings predicted by warm-fluid theory are reliable in regimes where kinetic effects are expected to be important.

In addition, nonlinear solutions of the three-wave equations can be used to fit simulations, which allow the seed intensity to be much stronger such that stimulated scattering dominates spontaneous ones. The three-wave equations are nonlinear partial differential equations, which are difficult to solve analytically. Nevertheless, in some cases, analytical solutions may be obtained using the inverse scattering method \citep{Kaup79}, and the setup might be possible to realize in kinetic simulations. 
Moreover, numerical solutions of the three-wave equations can be used to fit kinetic simulations. Notice that analytical solutions of wave-like equations are typically given in integral form, which may not have simple closed-form expressions. Even for the linearized three-wave problem, the integrals need to be evaluated numerically, except for a few limiting cases. The necessity of numerical integration makes it appealing to bypass analytical steps to directly solve the nonlinear equations using numerical methods. Parameters like the coupling coefficient may be scanned such that numerical solutions of three-wave equations match kinetic simulations, which involve many more variables. Once the analytical formula for magnetized coupling coefficient is benchmarked, the much cheaper numerical solutions can be used in lieu of the much more expensive kinetic simulations to study three-wave interactions of interest.

Finally, interactions in higher spatial dimensions remain to be investigated, which involves additional physics that are important for magnetized inertial confinement fusion and can also be exploited as a technique to reduce numerical artifacts. 
For example, during crossbeam laser energy transfer, the pump and seed lasers propagate in different directions. When a background magnetic field is present in yet another direction, the interaction is intrinsically multidimensional. After Lorentz boost into a reference frame where the two lasers are counter propagating, the plasma has a background flow and the static magnetic field partially transforms into an electric field. However, the initial boundary value problem in a finite plasma slab is more difficult to transform. In this case, higher dimensional simulations are conceptually simpler even though they are more costly to run. 
Additionally, using extra spatial dimensions, pump and seed may become easier to separate. Notice that the reflected laser propagates at the specular angle. If the seed is at a different angle, then it can be separated from the reflected pump, which is difficult to achieve in one spatial dimension. Using the different propagation angles, contamination from spontaneous scattering may also be mitigated. In unmagnetized plasmas, spontaneous scattering is usually strongest in the backward direction. Although magnetization changes the preferred scattering angle, there is usually a direction where spontaneous scattering peaks. Away from that direction, the seed laser suffers less from spontaneous scattering, so weak resonances may then become extractible.

In summary, this paper derives analytical solutions of the linearized three-wave equations in one spatial dimension, and use the solutions to benchmark magnetized coupling coefficients predicted by warm-fluid theory. A protocol is developed to setup, calibrate, and post process PIC simulations, so that the simulation data can be fitted using analytical solution in the same setup. The rigorous protocol yields excellent agreement between theory and simulations in the backscattering geometry for a wide range of plasma temperature $T_0$, magnetic field strength $B_0$, and propagation angle $\theta_B$. Growth rates predicted by warm-fluid theory match fully kinetic simulations even when collisionless damping becomes significant. The protocol is applicable to strong resonances that suffer less from spontaneous scattering and pump-seed leakage. It remains to be investigated whether the agreement can be extended to weaker resonances and to higher temperatures.

This work was performed under the auspices of the U.S. Department of Energy by Lawrence Livermore National Laboratory under Contract DE-AC52-07NA27344 and is supported in part by LLNL-LDRD under Project Number 20-SI-002.

\section*{Data and code availability}
Data underlying figures in this paper are openly available in zenodo at \url{https://doi.org/10.5281/zenodo.7126300}, reference \citep{yuan_shi_2022}.
The PIC simulations are performed using version 4.17.10 of the EPOCH code at \url{https://github.com/Warwick-Plasma/epoch}, reference \citep{epoch}. 
The analytical three-wave coupling coefficients from warm-fluid theory are evaluated using version 2.2.0 of the Three-Wave-MATLAB code at \url{https://gitlab.com/seanYuanSHI/three-wave-matlab}, reference \citep{Shi22}.
The protocol for computing analytical solutions, setting up simulations, and analyzing data is summarized at \url{https://gitlab.com/seanYuanSHI/linear-three-wave-analysis/}, reference \citep{Shi22linear}.

\appendix
\section{Solution map of the coupled-mode equation}\label{appA}
In this appendix, we evaluate the Fourier integral in Eq.~(\ref{eq:Green}) and analyze properties of $\Phi$ in Eq.~(\ref{eq:Phix}), which gives the solution map of the linear differential operator $\mathsfbi{L}$ in Eq.~(\ref{eq:linear-three-wave}). We will focus on the nondegenerate case $\Delta v=v_2-v_3>0$.

The coupled-mode equation $\mathsfbi{L}\ubalpha=\mathbf{0}$ can be mapped to the standard wave equation. Using the transformed coordinates given by Eq.~(\ref{eq:COMframe}), the advective derivatives become $\partial_t+v_2\partial_x=\frac{1}{2}\Delta v(\partial_\tau+\partial_z)$ and $\partial_t+v_3\partial_x=\frac{1}{2}\Delta v(\partial_\tau-\partial_z)$. 
Moreover, after changing variable $\ubalpha=\rho\ubbeta$, where $\rho$ is given by Eq.~(\ref{eq:rho}), the damping terms are removed and
\begin{subeqnarray}
	\label{eq:betaEq}
	(\partial_\tau+\partial_z)\beta_2 & = & m\beta_3, \\
	(\partial_\tau-\partial_z)\beta_3 & = & m\beta_2.
\end{subeqnarray} 
Substituting one equation into the other, $\ubbeta$ satisfies $(\partial_\tau^2-\partial_z^2-m^2)\ubbeta=\mathbf{0}$, which is the Klein-Gordon equation with an imaginary mass term. Since the equation is invariant under Lorentz transformations, it can be solved in any inertial frame.

\begin{figure}
	\centerline{\includegraphics[width=0.35\textwidth]{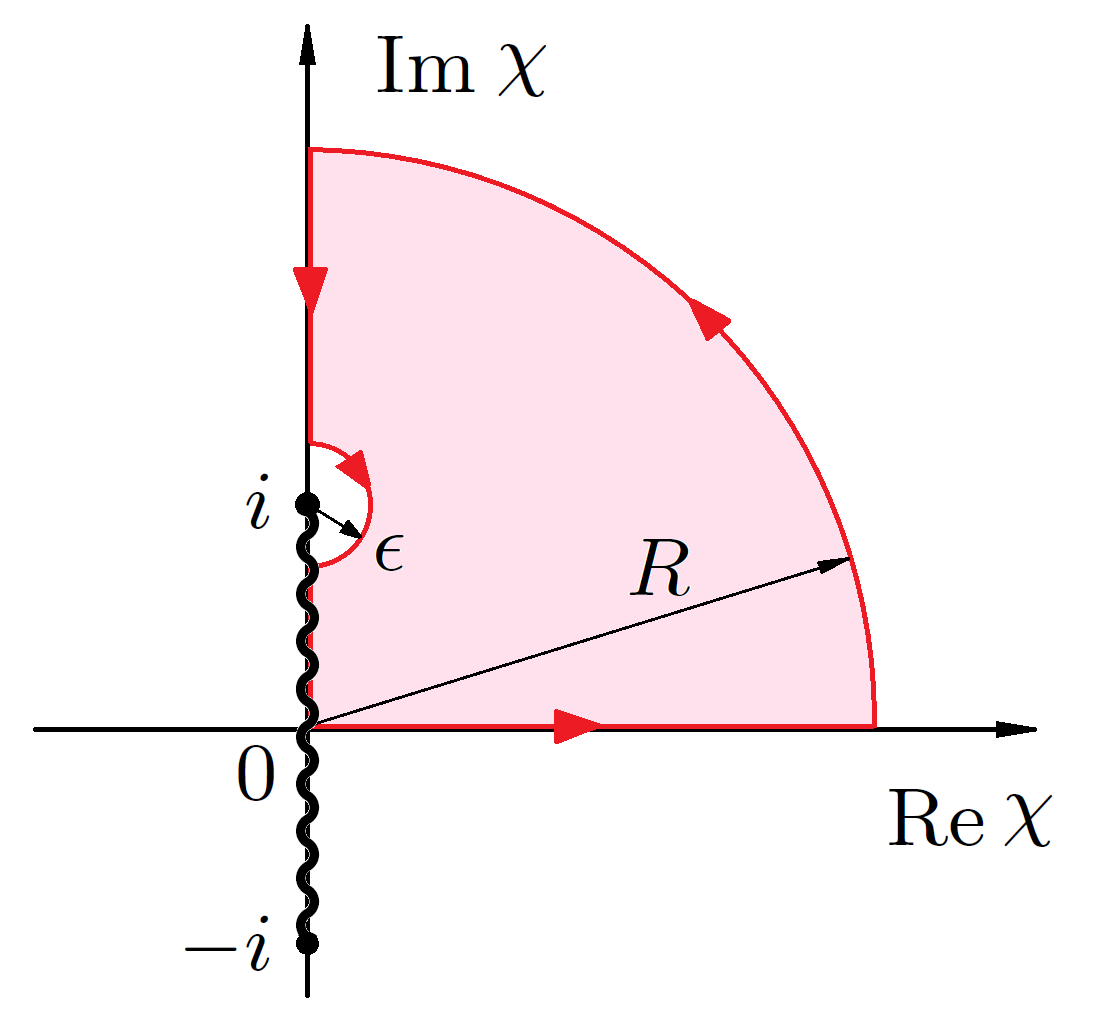}}% Images in 100% size
	\caption{Contour integral in complex $\chi$ plane. The integrands have a branch cut between $\chi=\pm i$, but is otherwise analytic. Since integrals along contours are zero, an integral along the real $\chi$ axis can be moved to the imaginary $\chi=i\eta$ axis by closing the contour at $R\rightarrow\infty$ and $\epsilon\rightarrow0$. When $\tau>0$, the closure is in the first quadrant as shown in this figure, whereas when $\tau<0$ the closure is in the fourth quadrant.}
	\label{fig:contour}
\end{figure}

Using Lorentz transformations, we can evaluate the integral in Eq.~(\ref{eq:Green}) in the complex $k'$ plane in three spacetime regions. First, when $(\tau,z)$ is a spacelike separation, namely $|\tau|<|z|$, we can Lorentz boost into a reference frame where $\tau'=0$. In this frame, the integrand is zero so $g=0$. Transforming back to the original frame
\begin{equation}
	\label{eq:g_spacelike}
	g(z,\tau)=0, \quad |\tau|<|z|,
\end{equation}
which enforces the causality that information outside the light cone does not contribute to the solution. 
Second, when $(\tau,z)$ is a timelike separation, namely $|\tau|>|z|$, we can Lorentz boost into a reference frame where $z'=0$. The direction of time is preserved if $\tau'=\pm\sqrt{\tau^2-z^2}$ takes the same sign as $\tau$. Then, the integrand becomes an even function and $g(\tau'/m)=\frac{1}{\pi}[\int_0^1 d\zeta \sinh(\tau'\sqrt{1-\zeta^2})/\sqrt{1-\zeta^2} + \int_1^{+\infty}d\zeta\sin(\tau'\sqrt{\zeta^2-1})/\sqrt{\zeta^2-1}]$. 
Changing variable to $\zeta=\sqrt{1+\chi^2}$, the second integral equals to the imaginary part of $\int_0^{+\infty}d\chi \, e^{i\tau'\chi}/\sqrt{1+\chi^2}$. 
Since the integrand is analytic away from the branch cut, the contour can be closed to become an integral along the imaginary $\chi=i\eta$ axis, as shown in figure~\ref{fig:contour}.
When $\tau'>0$, $e^{i\tau'\chi}$ decays when $\eta>0$, so the closure is in the first quadrant. 
For $\eta\in(1,\infty)$, the integral is real, so the only contribution comes from $\eta\in[0,1)$, which gives the imaginary part as $\int_0^1 d\eta\, e^{-\tau'\eta}/\sqrt{1-\eta^2}$. Substituting this into $g(\tau'/m)$ and changing variable $\zeta=\sqrt{1-\eta^2}$ in the first integral, we have $g(\tau'/m)=\frac{1}{\pi}\int_0^1 d\eta\, \cosh(\tau'\eta)/\sqrt{1-\eta^2}=\frac{1}{2}I_0(\tau')$, where $I_0$ is modified Bessel function \citep[Eq.~10.32.2]{DLMF}. 
Similarly, when $\tau'<0$, the contour integral is closed in the fourth quadrant, giving rise to an overall negative sign. 
In the original frame,
\begin{equation}
	\label{eq:g_timelike}
	g(z,\tau)=\frac{1}{2}\mathrm{sign}(\tau)I_0(m\sqrt{\tau^2-z^2}), \quad |\tau|>|z|.
\end{equation}
Finally, for lightlike separation $|\tau|=|z|$, $e^{ik'z}=\cos k'\tau\pm i\sin k'\tau$. Since $\hat{g}(k')$ is an even function, $g(z=\pm\tau/m)=\frac{1}{\pi}[\int_0^1 d\zeta\, \cos(\tau\zeta)\sinh(\tau\sqrt{1-\zeta^2})/\sqrt{1-\zeta^2} + \int_1^{+\infty} d\zeta\, \cos(\tau\zeta)\sin(\tau\sqrt{\zeta^2-1})/\sqrt{\zeta^2-1}]$. 
Changing variable to $\zeta=\sqrt{1+\chi^2}$, the second integral equals to the imaginary part of $\frac{1}{2}\int_0^{+\infty}d\chi \{\exp[i\tau(\sqrt{1+\chi^2}+\chi)] - \exp[i\tau(\sqrt{1+\chi^2}-\chi)] \}/\sqrt{1+\chi^2}$. 
Since $\sqrt{1+\chi^2}+\chi\simeq 2\chi$ when $\chi\rightarrow\infty$, the contour can be closed in first quadrant when $\tau>0$ to move the integral along $\chi=i\eta$ axis, as shown in figure~\ref{fig:contour}. For $\eta\in(1,+\infty)$, the integral is real, so the only contribution comes from $\eta\in[0,1)$, which gives the imaginary part as $\int_0^1d\eta\, \cos(\tau\sqrt{1-\eta^2})e^{-\tau\eta}/\sqrt{1-\eta^2}$. 
On the other hand, $\sqrt{1+\chi^2}-\chi\simeq 1/2\chi$ when $\chi\rightarrow\infty$, so closing the contour along $\chi=Re^{i\theta}$ gives a finite boundary term $i\int_0^{\pi/2} d\theta\, \chi \exp[i\tau(\sqrt{1+\chi^2}-\chi)]/\sqrt{1+\chi^2} \rightarrow i \int_0^{\pi/2} d\theta = i\pi/2$ when $R\rightarrow\infty$. The imaginary part of the second $\chi$ integral is therefore $-\pi/2+\int_0^1d\eta\, \cos(\tau\sqrt{1-\eta^2})e^{\tau\eta}/\sqrt{1-\eta^2}$.
Summing the two $\chi$ integrals leads to a cancellation with the first $\zeta=\sqrt{1-\eta^2}$ integral, leaving a constant $\frac{1}{4}$. 
Similarly, when $\tau<0$, the contour integration is closed in the fourth quadrant, giving rise to an overall negative sign. Therefore, along the light cone
\begin{equation}
	\label{eq:g_lightlike}
	g(z,\tau)=\frac{1}{4}\mathrm{sign}(\tau), \quad |\tau|=|z|.
\end{equation}
Using $I_0(0)=1$ and the definition of Heaviside step function with $\Theta(0)=\frac{1}{2}$, the inverse Fourier transform in all spacetime regions is combined into a single expression. 
%Notice that the argument of $\Theta$ can be any sign-preserving function of $\tau^2-z^2$. However, due to properties of the step function and its derivatives, the exact functional form is irrelevant and can be taken to be unity. 

The above $g$ satisfies the imaginary-mass Klein-Gordon equation, and consequently $\Phi$ is the solution map of the coupled-mode equation. Using $\partial_x\Theta(x)=\delta(x)$ and property of the $\delta$ function $\delta(f(x))=\sum_i\delta(x-x_i)/|f'(x_i)|$, where the summation is over all roots of $f(x)$, we have $\partial_\tau\Theta(\tau^2-z^2)=\frac{\tau}{|\tau|}[\delta(\tau-z)+\delta(\tau+z)]$ and $\partial_z\Theta(\tau^2-z^2)=\frac{\tau}{|\tau|}[\delta(z+\tau)-\delta(z-\tau)]$. 
Hence, derivatives of $g$ are $2\partial_\tau g = m^2|\tau|\Theta(\tau^2-z^2)I'_0(\xi)/\xi +\delta(\tau-z) + \delta(\tau+z)$, where $\xi=m\sqrt{\tau^2-z^2}$, and $2\partial_z g = -\mathrm{sign}(\tau)m^2z\Theta(\tau^2-z^2)I'_0(\xi)/\xi +\delta(z+\tau) - \delta(z-\tau)$.
Moreover, using $f(z,\tau)\partial_\tau\delta(\tau-z) = \frac{1}{2}\delta(\tau-z)(\partial_z-\partial_\tau)f(z,\tau)$ and similar properties of the $\delta$ function, we can calculate the second order partial derivatives. Since modified Bessel function satisfies $I_0''(\xi)+I'_0(\xi)/\xi=I_0(\xi)$, after canceling the $\delta$ functions,
\begin{equation}
	\label{eq:gGreen}
	(\partial_\tau^2-\partial_z^2-m^2) g(z,\tau)=0,
\end{equation}
which is consistent with $(\partial_\tau^2+k'^2-m^2)\hat{g}(k',\tau)=0$, where $\hat{g}(k',\tau)$ is in the integrand of Eq.~(\ref{eq:Green}). 
To see that $\Phi$ in Eq.~(\ref{eq:Phix}) is the solution map of $\mathsfbi{L}$ in Eq.~(\ref{eq:linear-three-wave}), commuting the damping factor $\rho$ in Eq.~(\ref{eq:rho}) with $\mathsfbi{L}$ gives the differential operator
\begin{equation}
	\label{eq:Drho}
	\mathsfbi{L}\rho = \frac{\Delta v}{2} \rho
	\left( \begin{array}{cc}
		\partial_\tau+\partial_z & -m \\
		-m & \partial_\tau-\partial_z
	\end{array} \right).
\end{equation}
Acting $\mathsfbi{L}$ on $\Phi$ leads to a diagonal differential operator on $g$ in Eq.~(\ref{eq:gGreen}) and gives $\mathsfbi{L}\Phi=\mathsfbi{0}$.
Finally, using earlier expressions of $\partial_\tau g$ and $\partial_z g$, Eq.~(\ref{eq:Phix}) is evaluated to
\begin{subeqnarray}
	\label{eq:PhixExplicit}
	&\frac{1}{\rho}\Phi(x,t) = \frac{\gamma_0}{\Delta v} \mathrm{sign}(t)\Theta[(v_2t\!-\!x)(x\!-\!v_3t)] \Phi_0(x,t)	
	 + 
    \left(\! \begin{array}{cc}
    	\delta(v_2t\!-\!x) & 0 \\
    	0 & \delta(x\!-\!v_3t)
    \end{array} \!\right),\quad \\
	&\Phi_0(x,t) =
	\left( \begin{array}{cc}
		I_1(\xi)\sqrt{\frac{x-v_3t}{v_2t-x}} & I_0(\xi) \\
		I_0(\xi) & I_1(\xi)\sqrt{\frac{v_2t-x}{x-v_3t}}
	\end{array} \right),
\end{subeqnarray}
where $\xi(x,t)=(2\gamma_0/\Delta v)\sqrt{(v_2t-x)(x-v_3t)}$. The kernel $\Phi_0$ is in the interior of the light cone, whose contribution vanishes when the coupling is zero. The $\delta$ function term is on the boundary of the light cone, which tracks the wave fronts as the waves advect.

%%%%%%%%%%%%%%%%%%%%%%%%%%%%%%%%%%%%%%%
%%%%%%%%%%%%%%%%%%%%%%%%%%%%%%%%%%%%%%%
\section{Well-posed boundary conditions}\label{appB}
In order for the backscattering problem $v_3<0$ to be well-posed, the boundary conditions need to satisfy constrains. 
In frequency domain, the constraints $(0, \tilde{l}_3(\omega))^\mathrm{T} =\tilde{\Psi}_0(\omega)\tilde{\mathbf{l}}(\omega)$ are degenerate, which give Eq.~(\ref{eq:l3w}) whose inverse is
\begin{equation}
	\label{eq:l2w}
	\tilde{l}_{2}(\omega)=\sqrt{\frac{|v_3|}{v_2}}\Big(\sigma+\sqrt{\sigma^2-1}\Big)\tilde{l}_{3}(\omega),
\end{equation}
where $\sigma=i\omega$. 
To find the relation between boundary conditions in time domain, take inverse Laplace transform $\mathcal{L}^{-1}[\tilde{p}(\sigma)](s)=\int_\mathcal{C} \frac{d\sigma}{2\pi i} e^{s\sigma}\tilde{p}(\sigma)$, where the contour $\mathcal{C}$ runs from $-i\infty$ to $+i\infty$ on the right-hand side of all poles. 
First, we need to find the inverse transforms of $\tilde{f}_\pm(\sigma)=\sigma\pm\sqrt{\sigma^2-1}$. Notice that $\tilde{f}'_{-}(\sigma)=-\tilde{I}_1(\sigma)$. In other words, $\tilde{f}'_{-}(\sigma)=\partial_\sigma\int_0^{+\infty} ds\, e^{-s\sigma}I_1(s)/s$, so $\mathcal{L}[I_1(s)/s](\sigma)=\int d\sigma\tilde{f}'_{-}(\sigma)=\tilde{f}_{-}(\sigma)+c_{-}$. To see that the integration constant $c_-$ is zero, notice that when $\sigma\rightarrow+\infty$, Laplace transforms approach zero, so is $\tilde{f}_{-}(\sigma)$. Therefore, the inverse transform is
\begin{equation}
	\label{eq:ILfn}
	\int_\mathcal{C} \frac{d\sigma}{2\pi i} e^{s\sigma} \big(\sigma-\sqrt{\sigma^2-1}\big) = \frac{I_1(s)}{s}.
\end{equation}
Using the above result and $\mathcal{L}^{-1}[\sigma](s)= \partial_s\int_{-\infty}^{+\infty} \frac{d\omega}{2\pi} e^{i\omega s}=\delta'(s)$, the inverse Laplace transform of $\tilde{f}_{+}(\sigma)=2\sigma-\tilde{f}_{-}(\sigma)$ is
\begin{equation}
	\label{eq:ILfp}
	\int_\mathcal{C} \frac{d\sigma}{2\pi i} e^{s\sigma} \big(\sigma+\sqrt{\sigma^2-1}\big) = 2\delta'(s)-\frac{I_1(s)}{s}.
\end{equation}
When $s>0$, the inverse transform of products is $\mathcal{L}^{-1}[\tilde{p}\tilde{q}](s)=\int_0^s ds'\,p(s')q(s-s')$. Therefore, the inverse Laplace transform of Eq.~(\ref{eq:l3w}) is given by Eq.~(\ref{eq:l3s}), and the inverse transform of Eq.~(\ref{eq:l2w}) is
\begin{equation}
	\label{eq:l2s}
	l_2(s) = \sqrt{\frac{|v_3|}{v_2}}\Big[2l'_3(s)-\int_0^s ds'\;\frac{I_1(s')}{s'} l_3(s-s') \Big].
\end{equation}

To show that Eqs.~(\ref{eq:l2s}) and (\ref{eq:l3s}) is an inversion pair, the following identity is needed. Substituting $l_2$ into $l_3$, the double integral of $l_3(s-s'-s'')$ can be simplified by changing the integration variables from $s'$ and $s''$ to $\chi=s'+s''$ and $\eta=s'-s''$, where the triangular domain of the double integral $ds'\,ds''=\frac{1}{2}d\eta\,d\chi$ is rotated to $\chi\in[0,s]$ and $\eta\in[-\chi,\chi]$. The inner $\eta$ integral is
\begin{equation}
	\label{eq:IdentityInverse}
	\int_{-\chi}^{\chi} d\eta\; \frac{I_1(\frac{\chi+\eta}{2})}{\chi+\eta} \frac{I_1(\frac{\chi-\eta}{2})}{\chi-\eta}
	= \frac{I_0(\chi)}{\chi}-\frac{2I_1(\chi)}{\chi^2}.
\end{equation}
To show the above identity, express $I_1(s)/s$ in terms of its inverse Laplace transform using Eq.~(\ref{eq:ILfn}), where $\frac{1}{2}(\chi\pm\eta)$ is paired with $\sigma_\pm$. Moving the $\sigma_\pm$ integrals outwards and computing the $\eta$ integral first leads to $\exp[\chi(\sigma_+ + \sigma_-)/2] \int_{-\chi}^{\chi} d\eta\, \exp[\eta(\sigma_+ - \sigma_-)/2]=2(e^{\chi\sigma_+}-e^{\chi\sigma_-})/(\sigma_+ - \sigma_-)$.
By the $\sigma_+\leftrightarrow\sigma_-$ symmetry of the outer integrals, it is sufficient to consider the $e^{\chi\sigma_+}$ term, which can be moved outside the $\sigma_-$ integral.
Closing the contour from right in the complex $\sigma_-$ plane, $\int_\mathcal{C_-} \frac{d\sigma_-}{2\pi i} [\sigma_- - (\sigma_-^2-1)^{1/2}]/(\sigma_+ - \sigma_-)= \frac{1}{2}[\sigma_+ - (\sigma_+^2-1)^{1/2}]$, where the factor $\frac{1}{2}$ comes from the fact that a pole is enclosed only when the contour $\mathcal{C_-}$ is on the left of $\mathcal{C_+}$.
To show that the remaining integral $\int_\mathcal{C_+} \frac{d\sigma_+}{4\pi i}e^{\chi\sigma_+} [\sigma_+ - (\sigma_+^2-1)^{1/2}]^2$ equals to the right-hand side of Eq.~(\ref{eq:IdentityInverse}), compute its Laplace transform $\tilde{F}(\sigma_+)=\int_0^{+\infty} d\chi e^{-\chi\sigma_+}[I_0(\chi)/\chi-2I_1(\chi)/\chi^2]$. 
Taking derivatives to remove denominators gives $\tilde{F}''(\sigma_+)=-\tilde{I}'_0(\sigma_+)-2\tilde{I}_1(\sigma_+)$,
which can be integrated twice to give $\tilde{F}(\sigma_+)=\sigma_+^2-\sigma_+(\sigma_+^2-1)^{1/2}+c_1\sigma_+ + c_2$ using $\tilde{I}_0$ and $\tilde{I}_1$ found before Eq.~(\ref{eq:l3w}). 
The integration constants $c_1=0$ and $c_2=-\frac{1}{2}$ are determined using the condition $\tilde{F}(+\infty) = 0$, which gives $\tilde{F}(\sigma_+)=\frac{1}{2}[\sigma_+ - (\sigma_+^2-1)^{1/2}]^2$ as desired.
Finally, substituting Eq.~(\ref{eq:IdentityInverse}) into the $l_3(s-s'-s'')$ integral partially cancels with the $l_3'(s-s')$ term after integration by part, and what remains are the boundary terms $l_3(s)-2l_3(0)I_1(s)/s$. The second term vanishes because $l_3(0)=0$, which is a consequence of Eq.~(\ref{eq:l3s}). We have thus shown that $l_3[l_2]=l_3$. Similarly, substituting $l_3$ into $l_2$ and using Eq.~(\ref{eq:IdentityInverse}), the converse can also be shown.

Verifying that $l_2$ and $l_3$ satisfy the constraints in time domain requires the following identity. For the first constraint $0=\int_0^s ds'[I_1(s')l_2(s-s')-\sqrt{|v_3|/v_2}I_0(s')l_3(s-s')]$, express $l_3$ in terms of $l_2$ using Eq.~(\ref{eq:l3s}) and change the double integral of $l_2(s-s'-s'')$ from $s'$ and $s''$ to $\chi=s'+s''$ and $\eta=s'-s''$. The constraint is satisfied because
\begin{equation}
	\label{eq:IdentityConstraint}
	\int_{-\chi}^{\chi} d\eta\; I_0\big(\frac{\chi+\eta}{2}\big) \frac{I_1(\frac{\chi-\eta}{2})}{\chi-\eta}
	= I_1(\chi).
\end{equation}
The proof of this identity is similar to Eq.~(\ref{eq:IdentityInverse}), where $I_0(s)$ and $I_1(s)/s$ are expressed in terms of their inverse Laplace transforms and the $\eta$ integral is performed before closing the $\sigma_\pm$ contours. 
Finally, for the second constraint $2l_3(s)=\int_0^s ds'[\sqrt{v_2/|v_3|}I_0(s')l_2(s-s') -I_1(s')l_3(s-s')]$, express $l_2$ in terms of $l_3$ using Eq.~(\ref{eq:l2s}). After integration by part for the $l'_3(s-s')$ term, which yields a boundary term and an integral that cancels the remaining terms due to Eq.~(\ref{eq:IdentityConstraint}), it is easy to see that the constraint is also satisfied.

%%%%%%%%%%%%%%%%%%%%%%%%%%%%%%%%%%%%%%%
%%%%%%%%%%%%%%%%%%%%%%%%%%%%%%%%%%%%%%%
\section{Properties of kernel functions}\label{appC}
To verify that Eq.~(\ref{eq:alphaStep}) satisfies boundary conditions, special values of $M_2$ and $M_3$ are needed. At the boundary, $\vartheta=0$ and $M_2(\varphi,0)=\int_0^1dr\, I_0(r\varphi)I_1(\varphi(1-r))/(1-r)$. The integral recovers Eq.~(\ref{eq:IdentityConstraint}) after changing variable to $\varphi'=(2r-1)\varphi$, so
\begin{equation}
	\label{eq:M2}
	M_2(\varphi, 0) = I_1(\varphi).
\end{equation}
Hence, $D_2(\varphi, 0)=0$ and $\alpha_2(x=0, t)=h_0$ satisfies the boundary condition. Similarly, changing the integration variable to $\varphi'$ gives
\begin{equation}
	\label{eq:M3}
	M_3(\varphi,0) = \int_{-\varphi}^{\varphi} d\varphi'\; I_1\big(\frac{\varphi+\varphi'}{2}\big)\frac{I_1(\frac{\varphi-\varphi'}{2})}{\varphi-\varphi'}=I_0(\varphi)-2\frac{I_1(\varphi)}{\varphi}.
\end{equation}
The proof of this identity is similar to Eqs.~(\ref{eq:IdentityInverse}) and (\ref{eq:IdentityConstraint}). 
Substituting $D_3(\varphi,0)=2I_1(\varphi)/\varphi$ into Eq.~(\ref{eq:alphaStep}) recovers $\alpha_3(x=0,t)=h_3(t)$. 
The other special values are at $\varphi\rightarrow0$. Using $I_1(\varphi)\simeq \varphi/2$, $M_2(0,\vartheta)=M_3(0,\vartheta)=0$, $D_2(0,\vartheta)=\vartheta$, and $D_3(0,\vartheta)=1$.

To verify that Eq.~(\ref{eq:alphaStep}) solves $\mathsfbi{L}\ubalpha=\mathbf{0}$, differential properties of the kernel functions are needed. Using $d_y\int^{b(y)}_{a(y)} dx\, f(x,y)=\int^{b}_{a} dx\, \partial_yf(x,y) + b'(y)f(b,y) - a'(y)f(a,y)$ where $a$ and $b$ are evaluated at $y$, the derivatives $\partial_t\alpha_2$ and $\partial_x\alpha_2$ can be easily computed, where the later involves $\partial_\vartheta D_2(\varphi,\vartheta)$. Using properties of modified Bessel functions,
\begin{subeqnarray}
	\label{eq:Derivative2}
	\partial_\vartheta M_2(\varphi,\vartheta) &=& M_3(\varphi,\vartheta),\\
	\partial_\vartheta D_2(\varphi,\vartheta) &=& D_3(\varphi,\vartheta),
\end{subeqnarray}
which ensures $(\partial_t+v_2\partial_x+\mu_2)\alpha_2=\gamma_0\alpha_3$ is satisfied by Eq.~(\ref{eq:alphaStep}). 
To show that $(\partial_t+v_3\partial_x+\mu_3)\alpha_3=\gamma_0\alpha_2$ is also satisfied, notice that $(v_3\partial_x+\mu_3) e^{-\mu_2x/v_2}$ gives a term proportional to $\int_0^{\gamma(t-x/v_2)}d\varphi\, D_3(\varphi,\vartheta) \partial_\varphi e^{-\varphi\gamma_a/\gamma_0}$, which can be computed using integration by part.
Using the special value $D_3(0,\vartheta)=1$ and identities
\begin{equation}
	\label{eq:Derivative3}
	(2\partial_\varphi-\partial_\vartheta)D_3(\varphi,\vartheta) - D_2(\varphi,\vartheta) = M_2(\varphi,\vartheta)-(2\partial_\varphi-\partial_\vartheta)M_3(\varphi,\vartheta) = 0,
\end{equation}
the $\alpha_3$ equation is straightforward to verify. 
To show the first equality in Eq.~(\ref{eq:Derivative3}), use the definitions of $D_2$ and $D_3$ before Eq.~(\ref{eq:Mfunction}) and $(2\partial_\varphi-\partial_\vartheta)I_0 = \sqrt{1+2\vartheta/\varphi}I_1$.
To show the second equality, $(2\partial_\varphi-\partial_\vartheta)M_3(\varphi,\vartheta) = \int_0^1dr\{ [I'_0(2\partial_\varphi\partial_\vartheta\xi-\partial_\vartheta^2\xi) + I''_0(2\partial_\varphi\xi-\partial_\vartheta\xi)\partial_\vartheta\xi]I_1/(1-r) + 2I'_0I'_1\partial_\vartheta\xi\}$, where the argument of $I_0$ is $\xi=\sqrt{r^2\varphi^2+2r\varphi\vartheta}$ and the argument of $I_1$ is $\varphi(1-r)$. 
Using derivatives of $\xi$ and the differential equation for $I_0$, the two terms in the square bracket combine to $r^2\varphi[(2r-1)\varphi+2\vartheta]I_0/\xi^2-2r^2\varphi^2(r-1)I'_0/\xi^3$. 
The remaining term is $\int_0^1dr\, I'_0I'_1\partial_\vartheta\xi = -\int_0^1dr\, (rI'_0/\xi)\partial_r I_1 =\int_0^1dr\, I_1\partial_r(rI'_0/\xi)$, because the boundary terms are zero. This term equals to $\int_0^1dr\, I_1[(1-r\varphi\vartheta/\xi^2)I_0-r^2\varphi^2I'_0/\xi^3]$.
Summing all contributions, the $I'_0$ terms cancel and $(2\partial_\varphi-\partial_\vartheta)M_3(\varphi,\vartheta) = \int_0^1dr\,cI_0I_1/(1-r)$, where $c=r^2\varphi[(2r-1)\varphi+2\vartheta]/\xi^2+2(1-r)(1-r\varphi\vartheta/\xi^2)=1$. Using the definition of $M_2$ in Eq.~(\ref{eq:Mfunction}), the second equality in Eq.~(\ref{eq:Derivative3}) is thus proved.

Finally, let us analyze the limit $t\rightarrow+\infty$ for two special cases. First, at the boundary, $\alpha_3(x=0,t)=h_0\sqrt{v_2/|v_3|} \Delta(\gamma t, \varsigma)$, where $\Delta(s, \varsigma)=\int_0^{s} ds'\, e^{-\varsigma s'} I_1(s')/s'$
and $\varsigma=\gamma_a/\gamma_0$ measures damping relative to growth. When $\gamma_0>\gamma_a$ is above the absolute instability threshold, the integral diverges and $\Delta(+\infty, \varsigma<1)=+\infty$. On the other hand, when $\gamma_0\le\gamma_a$, the integral converges and is given by Eq.~(\ref{eq:ILfn}) as $\Delta(+\infty, \varsigma\ge1)=\varsigma-\sqrt{\varsigma^2-1}$. 
At the threshold $\gamma_0=\gamma_a$, using property of modified Bessel function \citep[Eq.~10.43.8]{DLMF}, the special value is $\Delta(s, \varsigma=1)=1-e^{-s}[I_0(s)+I_1(s)]\simeq 1-\sqrt{2/\pi s}$, so the steady state is approached only at an algebraic rate.
Second, at zero plasma wave velocity $v_3=0$, the solutions are given by Eq.~(\ref{eq:DeltaStep}). When $t\rightarrow+\infty$, the integrals are evaluated using property of modified Bessel function \citep[Eqs.~10.43.24]{DLMF} and $I_{1/2}(x)=\sqrt{2/\pi x}\sinh x$.
To see how fast the solutions approach steady states, consider the residual $\mathcal{R}=\int_\psi^{+\infty} d\xi \, e^{-\nu\xi^2}I_1(\xi)$. The residual diminishes exponentially as $e^{-\nu\psi^2}=e^{-\mu_3t_r}$ when $\psi\gg\xi_*$, where $t_r=t-x/v_2$ is the retarded time. The threshold $\xi_*$ maximizes the integrand of $\mathcal{R}$ and is the unique root of $I_0(\xi_*)/I_1(\xi_*)=1/\xi_*+2\nu\xi_*$.
When $\nu\gg1$, namely when the spatial gain is small, $\xi_*\simeq1/\sqrt{2\nu}$, so the growth saturates when $t_r\gg1/\mu_3$.
On the other hand, when $\nu\ll 1$, namely when the spatial gain is large, $\xi_*\simeq1/2\nu$, so the growth saturates when $t_r\gg\gamma_0^2x/\mu_3^2v_2$. 
To give a better approximation of $\mathcal{R}$ when $\nu\ll 1$, notice that $\xi>\psi\gg\xi_*\gg1$ so $I_1(\xi)\simeq e^\xi/\sqrt{2\pi\xi}$. Changing the integration variable gives $\mathcal{R}\simeq \frac{1}{2\sqrt{2\pi}} e^{1/4\nu}\int_{\zeta_0}^{+\infty} d\zeta\,e^{-\zeta}[\zeta(\frac{1}{2}+\sqrt{\nu\zeta})]^{-1/2}$, where $\zeta_0=\nu(\psi-1/2\nu)^2=\mu_3t_r-\sqrt{\mu_3t_r/\nu}+1/4\nu$. 
When $\nu\zeta_0\gg\frac{1}{4}$, which is equivalent to $t_r\gg4\gamma_0^2x/\mu_3^2v_2$, the $\frac{1}{2}$ term is negligible, and the integral is evaluated to $\mathcal{R}\simeq \frac{1}{2\sqrt{2\pi}} e^{1/4\nu} \nu^{-1/4}\Gamma(\frac{1}{4}, \zeta_0)$ using the incomplete gamma function $\Gamma(n,z)=\int_z^{+\infty} d\zeta \,\zeta^{n-1}e^{-\zeta}\simeq z^{n-1}e^{-z}$.

\bibliographystyle{jpp}
%\bibliography{literature}

\end{document}